\documentclass{pasa}

\usepackage{graphicx}
\usepackage{multirow}
\usepackage{caption, subcaption}
\usepackage{pdflscape}
\usepackage{amsmath}
\usepackage{tablefootnote}
\usepackage{siunitx}

\newcommand{\noopsort}[1]{}

\def\h {^{\mathrm{h}}}
\def\m {^{\mathrm{m}}}
\def\deg {^{\circ} }
\def\sqdeg {~deg$^2$}

\graphicspath{{/Users/franzen/Projects/RLF/}}

\title[The GLEAM 200~MHz Local Radio Luminosity Function]{The GLEAM 200~MHz Local Radio Luminosity Function for AGN and Star-forming Galaxies}

\author[Franzen et al.]{T.~M.~O.~Franzen$^{1,2}$\thanks{Email: franzen@astron.nl} ,
N.~Seymour$^{2}$,
E.~M.~Sadler$^{3,4}$,
T.~Mauch$^{5}$,
S.~V.~White$^{6}$,
C.~A.~Jackson$^{1}$,
R.~Chhetri$^{2}$,
B.~Quici$^{2}$,
M.~E.~Bell$^{4}$,
J.~R.~Callingham$^{1}$,
K.~S.~Dwarakanath$^{7}$,
B.~For$^{8}$,
B.~M.~Gaensler$^{9}$,
P.~J.~Hancock$^{2}$,
L.~Hindson$^{10}$,
N.~Hurley-Walker$^{2}$,
M.~Johnston-Hollitt$^{2,11}$,
A.~D.~Kapi\'{n}ska$^{12}$,
E.~Lenc$^{4}$,
B.~McKinley$^{2}$,
J.~Morgan$^{2}$,
A.~R.~Offringa$^{1}$,
P.~Procopio$^{13}$,
L.~Staveley-Smith$^{8}$,
R.~B.~Wayth$^{2}$,
C.~Wu$^{8}$
and Q.~Zheng$^{10}$
\affil{$^1$ASTRON: the Netherlands Institute for Radio Astronomy, PO Box 2, 7990 AA, Dwingeloo, The Netherlands}
\affil{$^2$International Centre for Radio Astronomy Research, Curtin University, Bentley, WA 6102, Australia}
\affil{$^3$Sydney Institute for Astronomy, School of Physics, The University of Sydney, NSW 2006, Australia}
\affil{$^4$CSIRO Astronomy and Space Science (CASS), PO Box 76, Epping, NSW 1710, Australia}
\affil{$^5$South African Radio Astronomy Observatory, 2 Fir Street, Black River Park, Observatory 7925, South Africa}
\affil{$^6$Department of Physics and Electronics, Rhodes University, PO Box 94, Grahamstown, 6140, South Africa}
\affil{$^7$Raman Research Institute, Bangalore 560080, India}
\affil{$^8$International Centre for Radio Astronomy Research, University of Western Australia, Crawley 6009, Australia}
\affil{$^9$Dunlap Institute for Astronomy and Astrophysics, University of Toronto, ON, M5S 3H4, Canada}
\affil{$^{10}$School of Chemical \& Physical Sciences, Victoria University of Wellington, Wellington 6140, New Zealand}
\affil{$^{11}$Curtin Institute for Computation, Curtin University, GPO Box U1987, Perth, WA 6845, Australia}
\affil{$^{12}$National Radio Astronomy Observatory, 1003 Lopezville Rd, Socorro, NM 87801, USA}
\affil{$^{13}$School of Physics, The University of Melbourne, Parkville, VIC 3010, Australia}
}

\jid{PASA}
\doi{10.1017/pas.\the\year.xxx}
\jyear{\the\year}

\usepackage{aas_macros}
\usepackage{hyperref} 
\hypersetup{colorlinks,citecolor=blue,linkcolor=blue,urlcolor=blue}

\hypersetup{draft}

\begin{document}

\begin{frontmatter}
\maketitle

\begin{abstract}
The GaLactic and Extragalactic All-sky Murchison Widefield Array (GLEAM) is a radio continuum survey at 76--227~MHz of the entire southern sky (Declination $<+30\deg$) with an angular resolution of $\approx 2$~arcmin. In this paper, we combine GLEAM data with optical spectroscopy from the 6dF Galaxy Survey to construct a sample of 1,590 local (median $z \approx 0.064$) radio sources with $S_{200\,\mathrm{MHz}} > 55$~mJy across an area of $\approx 16,700~\mathrm{deg}^{2}$. From the optical spectra, we identify the dominant physical process responsible for the radio emission from each galaxy: 73 per cent are fuelled by an active galactic nucleus (AGN) and 27 per cent by star formation. We present the local radio luminosity function for AGN and star-forming galaxies at 200~MHz and characterise the typical radio spectra of these two populations between 76~MHz and $\sim 1$~GHz. For the AGN, the median spectral index between 200~MHz and $\sim 1$~GHz, $\alpha_{\mathrm{high}}$, is $-0.600 \pm 0.010$ (where $S \propto \nu^{\alpha}$) and the median spectral index within the GLEAM band, $\alpha_{\mathrm{low}}$, is $-0.704 \pm 0.011$. For the star-forming galaxies, the median value of $\alpha_{\mathrm{high}}$ is $-0.650 \pm 0.010$ and the median value of $\alpha_{\mathrm{low}}$ is $-0.596 \pm 0.015$. Among the AGN population, flat-spectrum sources are more common at lower radio luminosity, suggesting the existence of a significant population of weak radio AGN that remain core-dominated even at low frequencies. However, around 4 per cent of local radio AGN have ultra-steep radio spectra at low frequencies ($\alpha_{\mathrm{low}} < -1.2$). These ultra-steep-spectrum sources span a wide range in radio luminosity, and further work is needed to clarify their nature.
\end{abstract}

\end{frontmatter}

\section{Introduction}\label{Introduction}

The local radio luminosity function (RLF) is a measure of the variation in the global average space density of radio sources with luminosity at the present epoch. It provides an essential benchmark from which to analyse the cosmic evolution of active galactic nuclei (AGN) and star-forming (SF) galaxies to high redshift \citep[see e.g.][]{condon1989} and is a key observable in extragalactic radio source population models.

The radio source population in the local universe has to date been best studied at 1.4~GHz by combining large-area radio continuum and optical redshift surveys \citep{condon2002,sadler2002,mauch2007,best2012}. Members of the two source classes (AGN and SF) can usually be distinguished using optical spectra. The local RLFs of these two source populations cross over at a 1.4~GHz radio luminosity of $\approx 10^{23}~\mathrm{W}~\mathrm{Hz}^{-1}$, SF galaxies dominating the population of radio sources below this power and radio-loud AGN above it.

In constraining radio source population models, it is important to use samples selected at a wide range of frequencies because of the changing nature of the sources contributing to the local RLF with frequency.

The radio emission from radio-loud AGN at low frequencies ($\lesssim 200$~MHz) mainly arises from the radio lobes rather than the radio core, hotspots and jets that dominate the emission of sources at high radio frequencies. Since the radio lobes are not subject to relativistic beaming or `Doppler boosting'  \citep[see e.g.][]{blandford1979}, low-frequency surveys allow radio-loud AGN to be selected independently of the orientation of the jet axis. Low-frequency surveys can also reveal past activity that is not evident at higher frequencies: large-scale, low-frequency radio emission evolves on relatively long timescales, providing a measure of the long term jet activity. High-frequency radio emission generally evolves on much shorter timescales, providing a better measure of current jet activity \citep[see e.g.][]{sadler2006,hurleywalker2015}.

Table~\ref{tab:sample_comparison} compares some of the most recent radio-optical samples used to derive the local RLF. At high frequencies, \cite{sadler2013} measured the local RLF at 20~GHz by matching radio sources from the Australia Telescope 20~GHz (AT20G) survey \citep{murphy2010} with nearby galaxies from the Third Data Release of the 6dF Galaxy Survey \citep[6dFGS DR3;][]{jones2009}. Although their sample contained some FRI and FRII radio galaxies, it was dominated by compact \citep[FR-0;][]{ghisellini2011} radio AGN without any extended radio emission apparent at lower frequencies. The observed properties of these compact 20\,GHz sources are consistent with them being a mixed population including young Compact Steep-Spectrum (CSS) and Gigahertz-Peaked Spectrum (GPS) radio galaxies \citep{odea1998}. The AT20G catalogue included very few galaxies where the radio emission arose mainly from star-formation processes, so the 20\,GHz RLF was only measured for the AGN population. 

At low frequencies, \cite{prescott2016} used data from the Giant Metrewave Radio Telescope (GMRT) to measure the 325\,MHz local RLF for a small sample of nearby ($z<0.25$) radio-detected AGN and star-forming galaxies in a 138\,deg$^2$ area covered by the Galaxy and Mass Assembly \citep[GAMA;][]{driver2009} survey. More recently, \cite{sabater2019} derived the local ($z < 0.3$) RLF at 150~MHz for radio AGN and SF galaxies separately for a much larger sample by cross-matching the first data release of the LOFAR Two-metre Sky Survey \citep[LoTSS DR1;][]{shimwell2019} with the Sloan Digital Sky Survey \citep[SDSS;][]{york2000,stoughton2002} main galaxy spectroscopic sample. Their radio AGN luminosity function is in good agreement with previous determinations of the local RLF at 1.4~GHz assuming a spectral index $\alpha = -0.7$ (where $S \propto \nu^{\alpha}$).

In this paper, we study the local radio source population at 200~MHz over most of the southern sky ($16,679~\mathrm{deg}^{2}$) by combining data from the GaLactic and Extragalactic All-sky MWA (GLEAM) survey \citep{wayth2015} with 6dFGS DR3. GLEAM is a wide-area radio continuum survey at 76--227~MHz with an angular resolution of $\approx 2$~arcmin, conducted with the Murchison Widefield Array \citep[MWA;][]{tingay2013}. The large fractional bandwidth of the MWA makes it possible to measure in-band spectral indices for the vast majority of the sources detected in GLEAM. In addition, the compact antenna layout of the MWA gives it extremely high surface brightness sensitivity, which is important for recovering extended radio emission in nearby galaxies.

We measure the local RLF for AGN and SF galaxies, and characterise the typical radio spectra of these two populations. Our local radio sample is far shallower than the LoTSS-SDSS sample by \cite{sabater2019} but covers a far wider area of sky, has a lower median redshift and contains a larger number of AGN at high radio luminosities ($P_{200\,\mathrm{MHz}} > 10^{24}~\mathrm{W}~\mathrm{Hz}^{-1}$). All sources in the sample have measured spectral indices between 200~MHz and $\sim 1$~GHz obtained using the NRAO VLA Sky Survey \citep[NVSS;][]{condon1998} at 1.4~GHz and the Sydney University Molonglo Sky Survey \citep[SUMSS;][]{mauch2003} at 843~MHz. In addition, the vast majority of the sources have GLEAM intra-band spectral indices, allowing their radio spectral properties to be studied in a systematic way and providing important information on the nature of nearby radio galaxies down to low luminosities.

The layout of the paper is as follows. The data used in this work are summarised in Section~\ref{Data}. We define the GLEAM-6dFGS sample in Section~\ref{Definition of the GLEAM-6dFGS sample}, and describe the radio properties of the sample in Section~\ref{Radio properties of the GLEAM-6dFGS sample}. In Section~\ref{The local radio luminosity function at 200 MHz}, the local RLF for AGN and SF galaxies is presented and compared with other measurements. Our results are summarised in Section~\ref{Summary and future work}. Throughout this paper we assume a Hubble constant of $70~\mathrm{km}~\mathrm{s}^{-1}~\mathrm{Mpc}^{-1}$ ($h = 0.70$), and matter and cosmological constant density parameters of $\Omega_{M} = 0.3$ and $\Omega_{\Lambda} = 0.7$. Right ascension is abbreviated as RA and declination is abbreviated as Dec.

\begin{table*}
\centering
\caption{Comparison of radio-selected samples used to derive the local RLF at frequencies between 150~MHz and 20~GHz. The samples extend to a maximum redshift of 0.2--0.3, corresponding to a look-back time of $\approx 2.4-3.4$~Gyr.}
\label{tab:sample_comparison}
\begin{tabular}{ccccccc}
\hline
 & & & \multicolumn{2}{c}{AGN} & \multicolumn{2}{c}{SF galaxy} \\
Sample & Magnitude \& & Area ($\mathrm{deg}^{2}$) & Number & Median $z$ & Number & Median $z$ \\
 & flux density limits &  &  &  &  &  \\
\hline
\vspace{0.25cm}
\begin{tabular}{@{}c@{}}GLEAM-6dFGS \\ (this paper)\end{tabular} & \begin{tabular}{@{}c@{}}$K < 12.65$ \\ $S_{200\, \mathrm{MHz}} > 55~\mathrm{mJy}^a$\end{tabular} & 16,679$^a$ & 1,157 & 0.081 & 427 & 0.015 \\
\vspace{0.25cm}
\begin{tabular}{@{}c@{}}LoTSS DR1-SDSS \\ \citep{sabater2019} \end{tabular} & \begin{tabular}{@{}c@{}}$14.5 < r < 17.77$ \\ $S_{150\, \mathrm{MHz}} \gtrsim 0.5~\mathrm{mJy}$\end{tabular} & 424 & 2,121 & 0.143 & 8,494 & 0.097 \\
\vspace{0.25cm}
\begin{tabular}{@{}c@{}}NVSS-2MASX \\ \citep{condon2019} \end{tabular} & \begin{tabular}{@{}c@{}}$K \leq 11.75$ \\ $S_{1.4\, \mathrm{GHz}} \geq 2.45~\mathrm{mJy}$\end{tabular} & 23,032 & 2,763 & 0.12 & 6,699 & 0.06 \\
\vspace{0.25cm}
\begin{tabular}{@{}c@{}}NVSS/FIRST-GAMA \\ \citep{pracy2016} \end{tabular} & \begin{tabular}{@{}c@{}}$m_{i} < 20.5$ \\ $S_{1.4\, \mathrm{GHz}} > 2.8~\mathrm{mJy}$\end{tabular} & $\approx 900$ & 1,692 & 0.15--0.20$^b$ & 527 & 0.05--0.10$^b$ \\
\vspace{0.25cm}
\begin{tabular}{@{}c@{}}GAMA-GMRT \\ \citep{prescott2016} \end{tabular} & \begin{tabular}{@{}c@{}}$r \leq 19.8$ \\ $S_{325\, \mathrm{MHz}} \gtrsim 5~\mathrm{mJy}$\end{tabular} & 138 & 134 & 0.14--0.16$^b$ & 38 & 0.06 \\
\vspace{0.25cm}
\begin{tabular}{@{}c@{}}AT20G-6dFGS \\ \citep{sadler2013} \end{tabular} & \begin{tabular}{@{}c@{}}$K < 12.75$ \\ $S_{20\, \mathrm{GHz}} > 50~\mathrm{mJy}$\end{tabular} & 16,980 & 202 & 0.058 & -- & -- \\
\vspace{0.25cm}
\begin{tabular}{@{}c@{}}NVSS-SDSS \\ \citep{best2012} \end{tabular} & \begin{tabular}{@{}c@{}}$14.5 < r < 17.77$ \\ $S_{1.4\, \mathrm{GHz}} > 5~\mathrm{mJy}$\end{tabular} & $\approx$ 7,100 & 7,302 & 0.16 & 1,866 & 0.055 \\
\begin{tabular}{@{}c@{}}NVSS-6dFGS \\ \citep{mauch2007} \end{tabular} & \begin{tabular}{@{}c@{}}$K < 12.75$ \\ $S_{1.4\, \mathrm{GHz}} > 2.8~\mathrm{mJy}$\end{tabular} & 7,076 & 2,661 & 0.073 & 4,006 & 0.035 \\
\hline
\end{tabular}
\begin{flushleft} \footnotesize{$^a$ For the GLEAM-6dFGS sample, there are two regions of different depths. The flux density limit in the deep region, and the total area of the deep and shallow regions, are quoted in the table. The flux density limit in the shallow region is 100~mJy. The area of the deep region is 5,113\sqdeg~and the area of the shallow region is 11,566\sqdeg. \newline $^b$ For the NVSS/FIRST-GAMA and GAMA-GMRT samples, the median redshifts are estimated from the redshift histograms.} \end{flushleft}
\end{table*}

\section{Data}\label{Data}

We describe the radio and optical data used to define the GLEAM-6dFGS sample.

\subsection{GLEAM Exgal and SGP catalogues and images (76--227~MH\lowercase{z})}\label{GLEAM Exgal/SGP catalogue and images}

We use the GLEAM Extragalactic \citep[Exgal;][]{hurleywalker2017} and South Galactic Pole \citep[SGP;][]{franzen2021} data releases as our low-frequency basis data. GLEAM Exgal is based on the first year (2013--2014) of GLEAM observations. It covers 24,831\sqdeg~at Dec $< +30\deg$, excluding the strip at Galactic latitude $|b| < 10\deg$ and a few regions around bright, complex sources such as the Magellanic Clouds. The GLEAM Exgal catalogue contains 307,455 source components with 20 separate flux density measurements between 76 and 227~MHz selected from a wide-band image centred at 200~MHz, with an angular resolution of $\approx 2$~arcmin. The typical rms noise in the wide-band image is $\approx 10$~mJy/beam. Spectral indices between 76 and 227~MHz were derived by fitting a power law to the sub-band flux densities.

GLEAM SGP is based on a subset of both years (2013--2015) of GLEAM observations. It covers an area of 5,113\sqdeg~surrounding the South Galactic Pole at $20\h 40\m < \mathrm{RA} < 05\h 04\m$ and $-48\deg < \mathrm{Dec} < -2\deg$. The GLEAM SGP catalogue contains 108,851 source components with 20 separate flux density measurements between 76 and 227~MHz selected from a wide-band image centred at 216~MHz, with an angular resolution of $\approx 2$~arcmin. The typical rms noise in the wide-band image is $\approx 5$~mJy/beam. This is still well above the classical confusion limit of $\approx 2$\,mJy/beam \citep{franzen2019}.

Spectral indices across the GLEAM band were derived by fitting a power law to the sub-band flux densities. The catalogue also contains integrated flux densities at 200~MHz, $S_{200}$, as in the GLEAM Exgal catalogue. If the GLEAM intra-band spectrum could be well-fitted by a power law, $S_{200}$ was derived from the power-law fit. Otherwise, $S_{200}$ was derived by extrapolating the 216~MHz integrated flux density from the wide-band image to 200~MHz assuming $\alpha = -0.8$.

\begin{figure*}
\begin{center}
\includegraphics[scale=0.58, angle=270, trim=0cm 0cm 6cm 0cm]{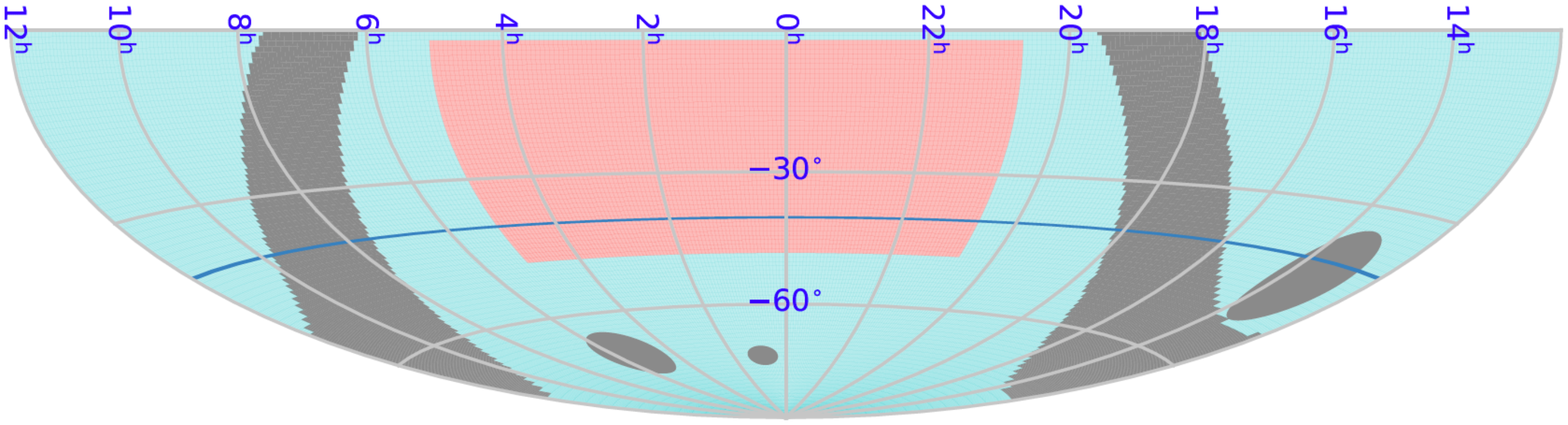}
\caption{The deep (both years: red) and shallow (year one only: cyan) regions of GLEAM used to define the GLEAM-6dFGS sample. The strip at Galactic latitude $|b| < 10\deg$ and a few regions surrounding Centaurus~A and the Magellanic Clouds, shown in grey, do not contain any GLEAM-6dFGS sources as they are not covered by GLEAM Exgal. The GLEAM-6dFGS sample is restricted to $\mathrm{Dec} < 0\deg$ due to the 6dFGS coverage. Flux densities at 1.4~GHz from NVSS are obtained above Dec $-39.5\deg$ (blue line) and flux densities at 843~MHz from SUMSS are obtained below Dec $-39.5\deg$. The Aitoff map projection is used.}
\label{fig:map_regions}
\end{center}
\end{figure*}

We refer to the area of sky covered by the GLEAM SGP catalogue, which is depicted by the red shading in Fig.~\ref{fig:map_regions}, as the `deep' region. We refer to the area of sky covered by the GLEAM Exgal catalogue at $\mathrm{Dec} < 0\deg$ and outside the deep region as the `shallow' region. The shallow region is depicted by the cyan shading in Fig.~\ref{fig:map_regions}. The deep and shallow regions are fully described in Table~\ref{tab:shallow_region}. In constructing the GLEAM-6dFGS sample, we use the GLEAM SGP catalogue in the deep region and the GLEAM Exgal catalogue in the shallow region.

\begin{table*}
\centering
\caption{Description of the deep and shallow regions of GLEAM used in the analysis of this paper.}
\begin{tabular}{ccc}
\hline
Description & Region & Area ($\mathrm{deg}^2$)\\
\hline
\textbf{Deep region} & $-48^{\circ}<\mathrm{Dec}<-2^{\circ}$ and $20^\mathrm{h}40^\mathrm{m}<\mathrm{RA}<05^\mathrm{h}04^\mathrm{m}$ & \textbf{5,113} \\
\hline
Galactic plane & Absolute Galactic latitude $<10^{\circ}$ & 3,578 \\
Centaurus~A & $13^\mathrm{h}25^\mathrm{m}28^\mathrm{s}$, $-43^{\circ}01^{\prime}09^{\prime\prime}$, $r=9^{\circ}$ & 254 \\
Large Magellanic Cloud & $05^\mathrm{h}23^\mathrm{m}35^\mathrm{s}$, $-69^{\circ}45^{\prime}22^{\prime\prime}$, $r=5.5^{\circ}$ & 95 \\
Small Magellanic Cloud & $00^\mathrm{h}52^\mathrm{m}38^\mathrm{s}$,  $-72^{\circ}48^{\prime}01^{\prime\prime}$, $r=2.5^{\circ}$ & 20 \\
Peeled sources$^\dagger$ & Radius of 10~arcmin & $< 1$ \\
\hline
\textbf{Shallow region} & \begin{tabular}{@{}c@{}}$\mathrm{Dec}<0^{\circ}$ excluding the deep region \\ and the regions listed in the middle rows\end{tabular} & \textbf{11,566} \\
\hline
\end{tabular}
\begin{flushleft}
\footnotesize{$^\dagger$ The peeled sources are Hydra A, Pictor A, Hercules A, Virgo A, Tau A, Cygnus A and Cassiopeia A; their positions are listed in \citet{hurleywalker2017}.}
\end{flushleft}
\label{tab:shallow_region}
\end{table*}

We use cutouts from the GLEAM Exgal and SGP wide-band images for visual inspection of the cross-identifications outlined in Section~\ref{Visual inspection}.

\subsection{SUMSS catalogue and images (843~MH\lowercase{z})}\label{SUMSS catalogue and images}

SUMSS covers the entire extragalactic sky ($|b|>10^\circ$) south of Dec $-30\deg$ at 843~MHz. It has a limiting peak brightness of 6~mJy/beam at $\mathrm{Dec} \leq -50\deg$ and 10~mJy/beam at $\mathrm{Dec} > -50\deg$. The angular resolution is $45\times45 \operatorname{cosec} |\mathrm{Dec}|~\mathrm{arcsec}^{2}$.

We use cutouts from SUMSS for the visual inspection at $\mathrm{Dec} < -39.5\deg$. If necessary, we use the SUMSS catalogue to correct the SUMSS flux density quoted in the NVSS/SUMSS-6dFGS catalogue described in Section~\ref{NVSS/SUMSS-6dFGS catalogue}.

\subsection{NVSS catalogue and images (1.4~GH\lowercase{z})}\label{NVSS catalogue and images}

NVSS covers the entire sky north of Dec $-40\deg$ at 1.4~GHz. It has a 5-$\sigma$ limit in peak brightness of $\approx 2.5$~mJy/beam and an angular resolution of 45~arcsec.

We use cutouts from NVSS for the visual inspection at $\mathrm{Dec} \geq -39.5\deg$. If necessary, we use the NVSS catalogue to correct the NVSS flux density quoted in the NVSS/SUMSS-6dFGS catalogue.

\subsection{The 6dFGS and the NVSS/SUMSS-6dFGS catalogue}\label{NVSS/SUMSS-6dFGS catalogue}

\subsubsection{The 6dFGS DR3 catalogue}
The 6dFGS DR3 catalogue \citep{jones2009} contains 125,071 galaxies, making near-complete samples with ($K$, $H$, $J$, $r_{F}$, $b_{J}$) $\leq$ 
(12.65, 12.95, 13.75, 15.60, 16.75); the median redshift is 0.053. It covers the whole sky south of Dec $0\deg$ and outside $10\deg$ of the Galactic plane. The redshift completeness of the catalogue is around 85\%, with some variation across the survey area.

\subsubsection{The NVSS/SUMSS-6dFGS catalogue}
As noted by \cite{white2020b}, many GLEAM sources have complex structure at low frequencies. Combined with the relatively low spatial resolution of the GLEAM images  ($\sim2-3$\,arcmin synthesised beam), this makes it challenging to identify the host galaxies of GLEAM sources in an automated way from low-frequency data alone. We therefore chose to use higher-frequency radio data from NVSS at 1.4~GHz and SUMSS at 843~MHz, which have higher spatial resolution than GLEAM, as a guide for cross-matching the 6dFGS and GLEAM catalogues as well as a source of radio spectral classifications for the GLEAM-6dFGS radio-matched galaxies.

\cite{mauch2007} used data from NVSS to identify 7,824 radio sources with galaxies brighter than K = 12.75 mag in the Second Incremental Data Release (DR2) of the 6dFGS. These authors later used the third and final 6dFGS data release (DR3) to compile a (currently unpublished) catalogue of NVSS-6dFGS and SUMSS-6dFGS cross-identifications at $\mathrm{Dec} \geq -39.5\deg$ and $\mathrm{Dec} < -39.5\deg$ respectively, which was used for this study.

The NVSS/SUMSS-6dFGS catalogue was generated using the same methodology described by \cite{mauch2007}, and contains $\approx$ 12,500 6dFGS-radio matches in total (representing around 10 per cent of the 125,071 galaxies in the 6dFGS DR3 catalogue).

A common method of determining the origin (star formation or jets originating from the central supermassive black hole) of the bulk of the radio emission in nearby galaxies is through their optical emission line properties. The optical spectrum of each galaxy in the NVSS/SUMSS-6dFGS catalogue was visually classified by Tom Mauch using the same method applied by \cite{mauch2007}. If the spectrum showed evidence of a galaxy with ongoing star formation, with strong narrow emission lines of H$\alpha$ and H$\beta$ typical of HII regions, the source was classified as a SF galaxy. Otherwise, the source was classified as an AGN. The AGN class was further divided into three sub-classes: (i) sources with pure absorption-line spectra, classed as `Aa'; (ii) sources with both absorption lines and weak, narrow emission lines characteristic of Low-Ionisation Nuclear Emission-line Regions (LINERs), classed as `Aae'; (iii) sources with strong nebular emission lines (stronger than any hydrogen Balmer emission lines), classed as `Ae'.

It is usually straightforward to tell the dominant physical process responsible for the radio emission from the optical spectra \citep{sadler1999,jackson2000}. The only significant complication is with the emission-line AGN (Ae and Aae objects) that make up 14 per cent of the GLEAM-6dFGS sample. These are a heterogeneous class of objects where both an AGN and star formation may contribute significantly to the radio emission, the dominant mechanism varying between sources \citep[e.g.][]{ching2017}. It is likely that some of the weaker radio sources in the Ae/Aae class arise mainly from star formation processes. We return to this point when deriving the local RLF for radio AGN and SF galaxies in section 5.2.

The NVSS/SUMSS-6dFGS catalogue includes the target name of the object from the 6dFGS data base; the position of the object and its total infrared $K$-band magnitude from the Two Micron All-Sky Survey Extended Source Catalogue \citep[2MASS XSC;][]{jarrett2000}, as listed in the 6dFGS database; the optical redshift and redshift quality flag, as listed in the 6dFGS catalogue; the spectral classification for galaxies with a good-quality 6dFGS spectrum from the visual inspection; and the integrated flux density of the radio counterpart in NVSS or SUMSS. 

Although NVSS and SUMSS flux densities are included in the NVSS/SUMSS-6dFGS catalogue, all the flux densities used in this paper were checked and re-derived from the original NVSS and SUMSS catalogues.

\subsection{The AllWISE catalogue in the mid-infrared}\label{The AllWISE catalogue in the mid-infrared}

The \textit{Wide-field Infrared Survey Explorer} \citep[WISE;][]{wright2010} covers the whole sky at 3.4 (W1), 4.6 (W2), 12 (W3) and 22 (W4)~$\mu \mathrm{m}$. We use the AllWISE data release catalogue by \cite{cutri2013} to characterise the GLEAM-6dFGS sources in the mid-infrared. The angular resolutions are 6.1, 6.4, 6.5 and 12.0~arcsec, and the flux sensitivities at 5$\sigma$ are 0.054, 0.071, 0.73 and 5.0~mJy/beam, in W1, W2, W3 and W4 respectively.

\subsection{Other radio and optical data}\label{Other radio and optical data}

The following data were also used, where available, for the visual inspection to confirm the cross-identification:
\begin{itemize}
  \item Cutouts from the alternative data release 1 of the TIFR GMRT Sky Survey \citep[TGSS ADR1;][]{intema2017}. The survey covers the entire sky north of Dec $-53\deg$ at 150~MHz to a typical rms sensitivity of 3.5~mJy/beam. The angular resolution is $25\times25 \operatorname{sec} (\mathrm{Dec} - 19\deg)~\mathrm{arcsec}^{2}$ at $\mathrm{Dec} < +19\deg$.
  \item Cutouts from the 1.4~GHz Faint Images of the Radio Sky at Twenty Centimetres \citep[FIRST;][]{becker1995} survey, which covers $\approx 10,575~\mathrm{deg}^{2}$ of the north Galactic cap. The typical sensitivity is 0.13~mJy/beam and the angular resolution 5~arcsec.
  \item Quick Look images from the first epoch of the VLA Sky Survey \citep[VLASS;][]{lacy2020}, which covers the entire sky at $\mathrm{Dec} > -40\deg$ at 2--4~GHz. The sensitivity is $\approx 120~\mu$Jy/beam and the angular resolution 2.5~arcsec.
  \item Optical blue ($B_{J}$-band) images from the SuperCOSMOS Sky Survey \citep{hambly2001} available over the entire southern sky to a typical AB depth of $B_{J} < 21$.
\end{itemize}

\section{Definition of the GLEAM-6\lowercase{d}FGS sample}\label{Definition of the GLEAM-6dFGS sample}

\subsection{Cross-matching GLEAM with 6dFGS, NVSS and SUMSS}\label{Cross-matching GLEAM with 6dFGS, NVSS and SUMSS}

\begin{figure}
\begin{center}
\includegraphics[scale=0.55, trim=0cm 0cm 0cm 0cm]{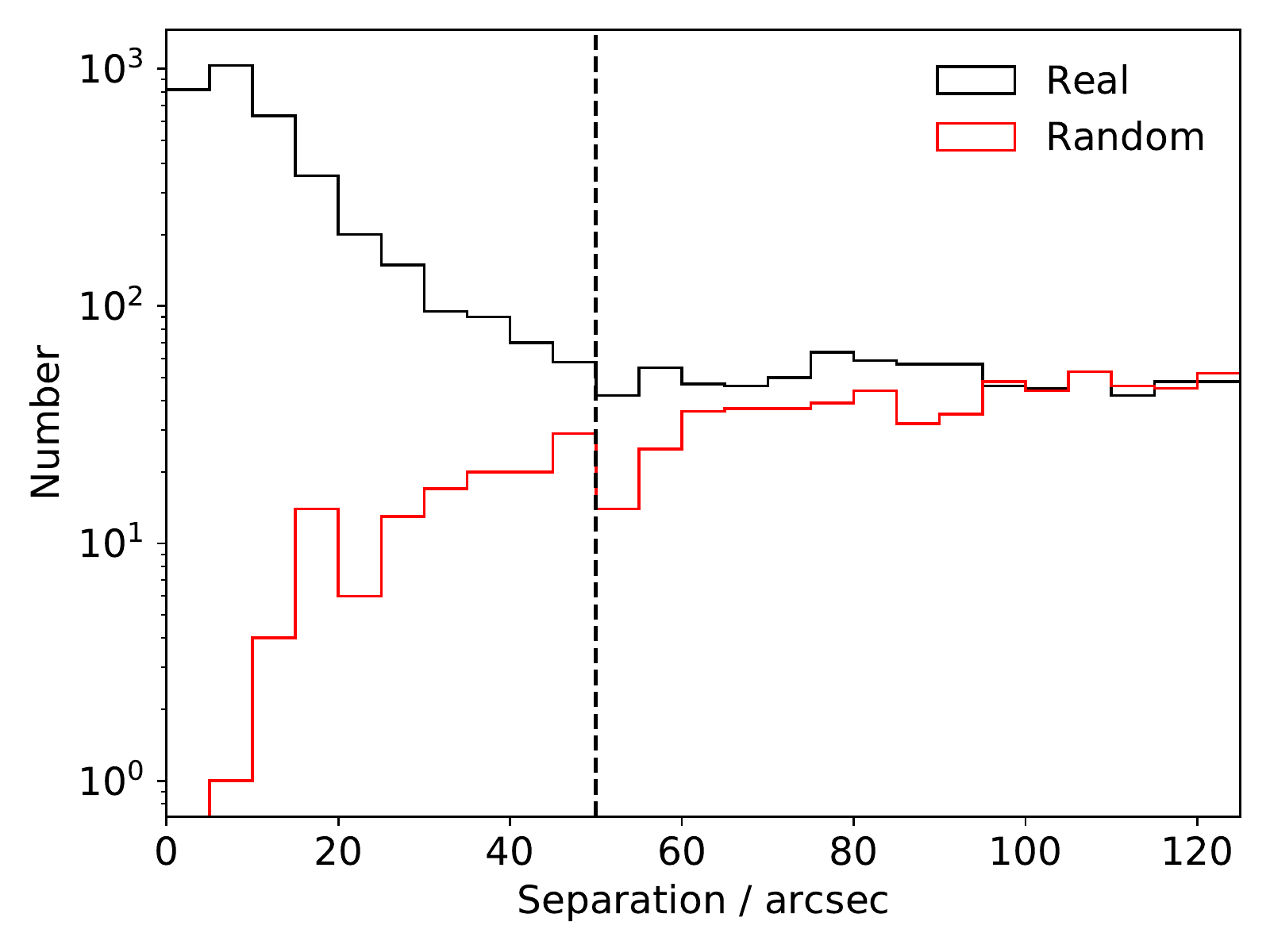}
\caption{Results of a simulation to determine the best search radius to automatically accept GLEAM counterparts to 6dFGS sources.
The black histogram shows the number of GLEAM sources with identified 6dFGS counterparts as a function of the angular separation between the GLEAM and 6dFGS positions. The red histogram shows the results obtained for a simulated catalogue generated by offsetting the GLEAM positions by 10~arcmin in Dec. The dashed vertical line at 50~arcsec marks the chosen search radius.}
\label{fig:monte_carlo_sim}
\end{center}
\end{figure}

We use the \textsc{topcat} software \citep{taylor2005} to cross-match the NVSS/SUMSS-6dFGS catalogue with the GLEAM SGP catalogue in the deep region and the GLEAM Exgal catalogue in the shallow region. We conduct a simulation to determine the best search radius to automatically accept GLEAM counterparts to 6dFGS galaxies. The black histogram in Fig.~\ref{fig:monte_carlo_sim} shows the number of GLEAM sources with identified 6dFGS counterparts as a function of angular separation between the GLEAM and 6dFGS positions. We generate a list of simulated sources by offsetting the GLEAM positions by 10~arcmin in Dec and cross-match the simulated catalogue with the NVSS/SUMSS-6dFGS catalogue. The red histogram shows the number of matches between the simulated catalogue and the NVSS/SUMSS-6dFGS catalogue as a function of angular separation. At a separation of 100~arcsec, the number of false matches is roughly equal to the number of real matches. Matches beyond this separation are therefore unlikely to be genuine. At a separation of 50~arcsec, roughly half the matches are false.

We conclude that a search radius of 50~arcsec provides a good compromise between completeness and reliability. Using this search radius, we obtain a total of 3,495 matches.

\subsubsection{Completeness and reliability of the cross-matched sample}\label{Completeness and reliability of cross-matched sample}

We use the results of the simulation to estimate the completeness and reliability of the matches. The simulation indicates that 124 random associations with radio sources are expected at separations less than 50~arcsec. This implies that, of the 3,495 accepted matches, $\approx 124$ are likely to be spurious, corresponding to a reliability of $1-\frac{124}{3495} = 96.5$ per cent. This reliability estimate should be taken as a lower limit because we subsequently remove a number of unreliable matches based on the visual inspection (see Section~\ref{Removing false GLEAM-6dFGS matches}).

All matches within the 50~arcsec search radius are accepted, hence the completeness is 100 per cent in this zone. We accept no radio source identifications at separations larger than 50~arcsec. However, based on the small excess of real over random matches with position offsets between 50 and 100~arcsec, we expect there to be 176 genuine associations in this zone. This implies that $\approx 176$ genuine matches are missing from our sample, corresponding to a completeness of $1-\frac{176}{3495} = 95.0$ per cent. The excess of real over random matches at separations greater than 50~arcsec could be, in part, due to clustering of optical galaxies (i.e. these excess matches could be associated with galaxy groups), as noted in previous studies \citep[e.g.][]{mauch2007,best2005}.

\subsection{Radio completeness}\label{Radio completeness}

We calculate the flux density levels at which the sample is close to 100 per cent complete in the shallow and deep regions. \cite{hurleywalker2017} created maps of the spatial variation of the completeness in GLEAM Exgal at 33 flux density levels spanning the range 25~mJy to 1~Jy. We use these maps to calculate the mean completeness at each flux density level within the shallow region. The mean completeness is found to be $\approx 90$ per cent at $S_{200\,\mathrm{MHz}} = 100$~mJy. The completeness is close to uniform except in a relatively small region of sky covering $\sim 500~\mathrm{deg}^{2}$ at $16\h < \mathrm{RA} < 20\h$ and $\mathrm{Dec} < -60\deg$, where it drops well below 90 per cent.

The GLEAM SGP catalogue is estimated to be 90 per cent complete at $S_{216\,\mathrm{MHz}} = 50$~mJy \citep{franzen2021}. The completeness is close to uniform across the entire region covered by the catalogue. The completeness limit at 200~MHz will depend on the distribution of spectral indices between 200 and 216~MHz. In estimating the completeness limit at 200~MHz, we take a conservative approach and assume that all sources have a spectral index of --1.2. Thus we set the completeness limit at 200~MHz to 55~mJy.

We note that genuine GLEAM-6dFGS matches which are too faint to appear in NVSS/SUMSS will be omitted from our sample. NVSS is close to 100 per cent complete at $S_{1400\,\mathrm{MHz}} = 3.4$~mJy. SUMSS has a completeness limit of $S_{843\,\mathrm{MHz}} = 10$~mJy at $\mathrm{Dec} \leq -50\deg$ and of $S_{843\,\mathrm{MHz}} = 18$~mJy at $\mathrm{Dec} > -50\deg$. Table~\ref{tab:area_alpha_lim} shows the limiting spectral index between 200 and 1400/843~MHz in five different regions of the GLEAM-6dFGS sample depending on the survey data available. The limiting spectral indices assume that the angular size of the source is not significantly larger than 45~arcsec, the angular resolution of NVSS and SUMSS.

\begin{table*}
\centering
\caption{The limiting spectral indices, $\alpha_{\mathrm{lim}}$, between 200 and 1400/843~MHz in five different regions of the GLEAM-6dFGS sample depending on the survey data available. Sources with $\alpha < \alpha_{\mathrm{lim}}$ may not appear in the sample. The final number of GLEAM-6dFGS sources in each region is given in the last column.}
\begin{tabular}{ccccccc}
\hline
& & \multicolumn{3}{c}{Completeness limit (mJy/beam)} & & \\
Region & Area ($\mathrm{deg}^2$) & $S_{200\,\mathrm{MHz}}$ & $S_{843\,\mathrm{MHz}}$ & $S_{1400\,\mathrm{MHz}}$ & $\alpha_{\mathrm{lim}}$ & $N$ \\
\hline
GLEAM deep \& $\mathrm{Dec} \geq -39.5\deg$ & 4,340 (26\%) & 55 & - & 3.4 & $-1.43$ & 536 \\ 
 GLEAM deep \& $\mathrm{Dec} < -39.5\deg$ & 773 (5\%) & 55 & 18 & - & $-0.78$ & 92 \\ 
 GLEAM shallow \& $\mathrm{Dec} \geq -39.5\deg$ & 6,879 (41\%) & 100 & - & 3.4 & $-1.74$ & 614 \\ 
 GLEAM shallow \& $-39.5\deg > \mathrm{Dec} \geq -50\deg$ & 1,163 (7\%) & 100 & 18 & - & $-1.19$ & 112 \\ 
 GLEAM shallow \& $\mathrm{Dec} < -50\deg$ & 3,524 (21\%) & 100 & 10 & - & $-1.60$ & 236 \\
\hline
\end{tabular}
\label{tab:area_alpha_lim}
\end{table*}

The highest limiting spectral index is --0.78 in the section of the deep region covered by SUMSS. The fraction of sources with spectral indices between 200~MHz and $\sim 1$~GHz, $\alpha_{\mathrm{high}}$, steeper than --0.78 is $\approx 21$ per cent based on the distribution of $\alpha_{\mathrm{high}}$ for the GLEAM-6dFGS sources (see Section~\ref{Radio spectra}). We therefore expect $\approx 21$ per cent of genuine GLEAM-6dFGS associations to be missing from the sample in this region close to the 55~mJy flux density cut, assuming that the source angular size is $\lesssim 45$~arcsec. However, this region only represents 5 per cent of the total area covered by the sample.

The limiting spectral indices in the other regions listed in Table~\ref{tab:area_alpha_lim} lie between --1.19 and --1.74. We can obtain an estimate of the fraction of sources with $\alpha < -1.2$ from Seymour et al., in preparation, who cross-matched GLEAM SGP with the third data release from the Australia Telescope Large Area Survey \citep[ATLAS DR3;][]{franzen2015}; ATLAS DR3 covers the \textit{Chandra} Deep Field South \citep[CDFS;][]{giacconi2001} and the European Large Area ISO Survey - South 1 \citep[ELAIS-S1;][]{oliver2000} to an rms depth of $\approx 15~\mu \mathrm{Jy/beam}$ at 1.4~GHz. Of the 134 GLEAM SGP sources lying in CDFS and ELAIS-S1, only four (3 per cent) were found to have $\alpha < -1.2$. From these spectral index limits we determine that the GLEAM-6dFGS sample is biased against diffuse radio sources with very steep spectra such as radio haloes and relics. Such sources may be detected by directly cross-matching the GLEAM catalogue with the 6dFGS catalogue, but this is beyond the scope of this paper. We do not expect any bias against this source class to significantly affect the overall completeness of the sample and the measured local RLF.

\subsection{Source filtering}\label{Source filtering}

The determination of the local RLF requires a sample which is complete to the limits of the radio and optical surveys from which it is derived. As described in Section~\ref{Radio completeness}, the shallow region is complete to $S_{200\,\mathrm{MHz}} = 100$~mJy and the deep region to $S_{200\,\mathrm{MHz}} = 55$~mJy. In order to form a sample which is complete in the radio, we discard all sources with $S_{200\,\mathrm{MHz}} < 100$~mJy and $S_{200\,\mathrm{MHz}} < 55$~mJy in the shallow and deep regions respectively. Since the 6dFGS DR3 catalogue is close to 100 per cent complete at $K < 12.65$~mag \citep{jones2009}, we discard sources with $K > 12.65$~mag. Finally, we remove sources with unreliable redshifts (redshift quality 1 or 2), those associated with Galactic stars (redshift quality 6) and those with $z < 0.001$, leaving a total of 1,688 sources.

\subsection{Visual inspection}\label{Visual inspection}

For all 1,688 selected sources, we visually inspect optical $J$-band images of size 15~arcmin from the SuperCOSMOS Sky Survey overlaid with radio contours from GLEAM and NVSS/SUMSS. Firstly, we identify and remove unreliable GLEAM-6dFGS matches. Secondly, we identify which NVSS/SUMSS components are associated with each source and, if necessary, correct the total NVSS/SUMSS flux density. Thirdly, we use the overlays to identify sources whose flux densities in GLEAM are significantly affected by confusion from adjacent sources and apply a deblending method to correct the GLEAM flux densities.

\subsubsection{Removing false GLEAM-6dFGS matches}\label{Removing false GLEAM-6dFGS matches}

The probability of a GLEAM-6dFGS match being genuine is dependent on the GLEAM-6dFGS offset. The simulations used to determine the best search radius to automatically accept GLEAM-6dFGS identifications (see Section~\ref{Cross-matching GLEAM with 6dFGS, NVSS and SUMSS}) indicate that the reliability of the cross-matches is $\approx 99$ per cent for position offsets less than 30~arcsec and $\approx 73$ per cent for position offsets between 30 and 50~arcsec. 

\begin{figure}
\begin{center}
\includegraphics[scale=0.35, angle=270, trim=0cm 2cm 0cm 2cm]{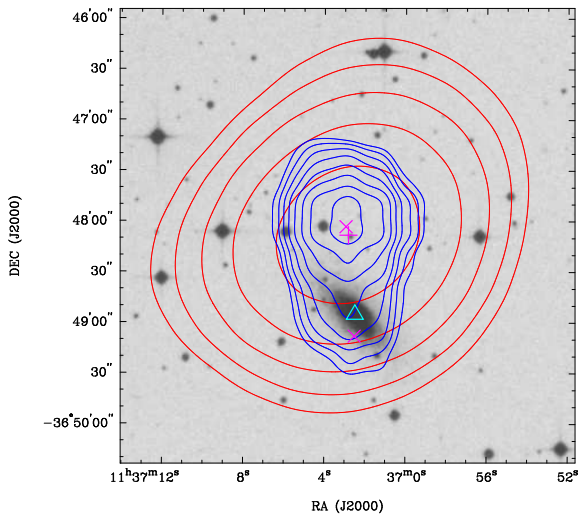}
\caption{Example overlay for a source with a relatively large GLEAM-6dFGS position offset (47.1~arcsec). Radio contours from GLEAM (200~MHz; red) and NVSS (1400~MHz; blue) are overlaid on the SuperCOSMOS J-band image (inverted greyscale). For each set of contours, the lowest contour is at the 3$\sigma$ level, where $\sigma$ is the local rms, with the number of $\sigma$ increasing by a factor of $\sqrt{2}$ with each subsequent contour. Catalogue positions from GLEAM (magenta plus signs) and NVSS (magenta crosses) are plotted. The cyan triangle shows the optical position of the source in 6dFGS. The large GLEAM-6dFGS position offset is the result of confusion with an adjacent source. The GLEAM-6dFGS match is not genuine and the source is removed from the sample.}
\label{fig:example_reject}
\end{center}
\end{figure}

We visually inspect the overlays of all sources, paying particular attention to sources with position offsets larger than 30~arcsec, and identify 22 false GLEAM-6dFGS matches. These sources are subsequently removed from the sample. The false GLEAM-6dFGS matches are typically the result of confusion with an adjacent source, displacing the position of the GLEAM component away from the 6dFGS position and boosting its flux density; an example of such a case is shown in Fig.~\ref{fig:example_reject}.

\subsubsection{Identifying multi-component sources}\label{Identifying multi-component sources}

We attempt to automatically identify multi-component sources in NVSS/SUMSS using the NVSS/SUMSS cutouts. The integration area of a source is taken to consist of all pixels in the NVSS/SUMSS image that are within the contiguous 3-$\sigma$ contour level, bounding the GLEAM position in question. All catalogued NVSS/SUMSS components located within the integration area are considered to be associated with the GLEAM component. We calculate the brightness-weighted centroid of the NVSS/SUMSS emission from the positions and flux densities of the associated NVSS/SUMSS components. The errors on the positions of the individual NVSS/SUMSS components are assumed to be correlated when calculating the centroid's position error. We also calculate the total flux density of the individual NVSS/SUMSS components. Again, we take a conservative approach and assume that the component flux density errors are correlated.

\begin{figure*}
\begin{center}
\includegraphics[scale=0.4, angle=270, trim=0cm 4cm 0cm 2cm]{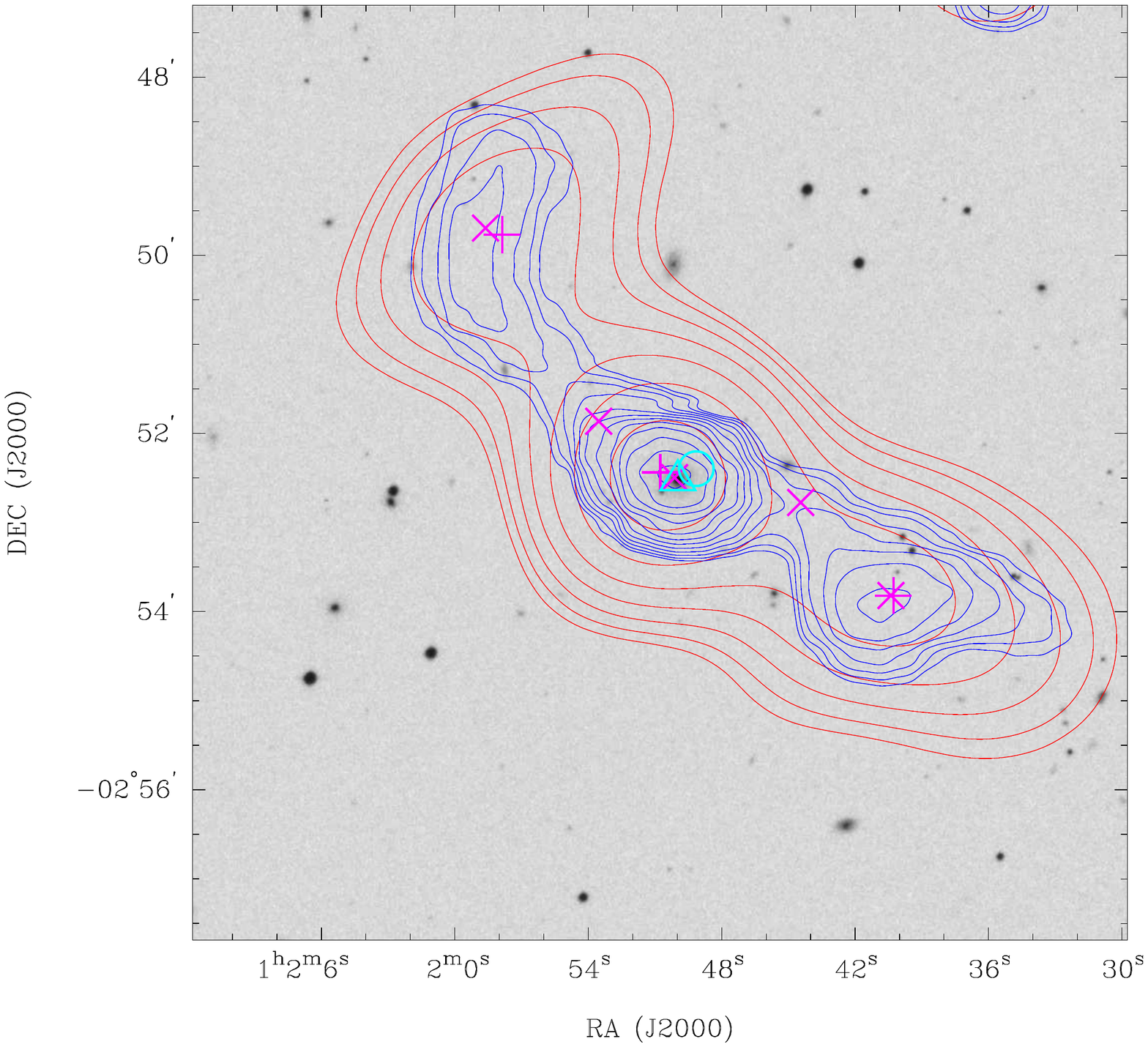}
\includegraphics[scale=0.4, angle=270, trim=0cm 4cm 0cm 4cm]{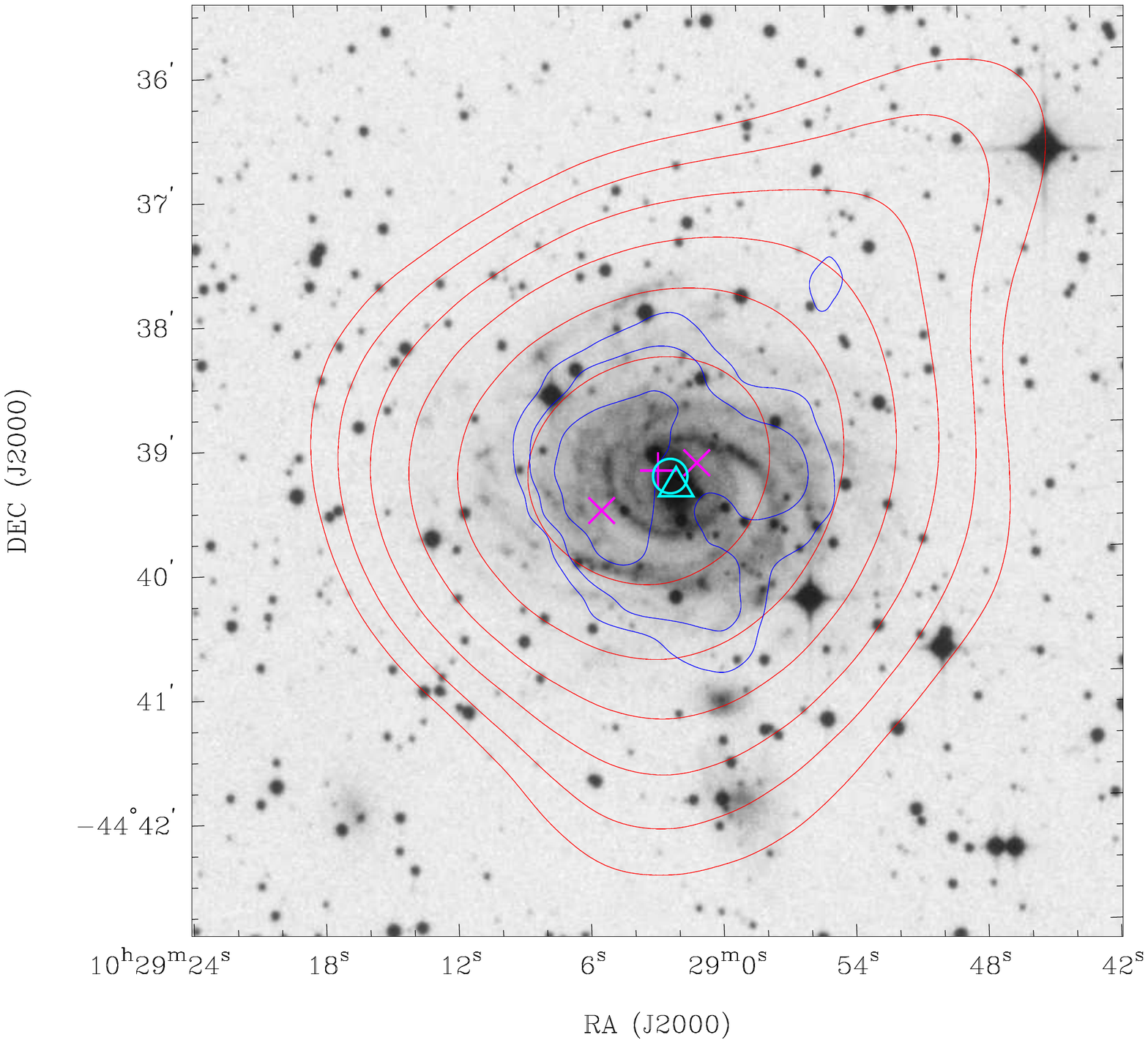}
\caption{Example overlays for an extended FRII at $z = 0.0706$ (GLEAM~J010150-025226; left) and a face-on spiral galaxy at $z = 0.0084$ (GLEAM~J102902-443916; right) in the GLEAM-6dFGS sample. Radio contours from GLEAM (200~MHz; red) and NVSS/SUMSS (1400/843~MHz; blue) are overlaid on the SuperCOSMOS J-band image (inverted greyscale). For each set of contours, the lowest contour is at the 3$\sigma$ level, where $\sigma$ is the local rms, with the number of $\sigma$ increasing by a factor of $\sqrt{2}$ with each subsequent contour. Catalogue positions from GLEAM (magenta plus signs) and NVSS or SUMSS (magenta crosses) are plotted. The brightness-weighted centroid position, calculated using the NVSS/SUMSS components, is indicated by a cyan circle. The cyan triangle shows the optical position of the source in 6dFGS.}
\label{fig:example_overlays}
\end{center}
\end{figure*}

This automated technique can fail to identify the correct NVSS/SUMSS components in the following two situations: (i) the NVSS/SUMSS components of a very extended source are well separated and there is no extended emission $> 3\sigma$ linking the components; (ii) unrelated point sources lie sufficiently close together on the sky such that they are located within the same integration area. In order to identify and correct these errors, we visually inspect all the overlays. Fig.~\ref{fig:example_overlays} shows example overlays for an extended AGN and SF galaxy where the NVSS/SUMSS components are correctly identified using the automated procedure.

\begin{figure*}
\begin{center}
\includegraphics[scale=0.4, angle=270, trim=0cm 4cm 0cm 2cm]{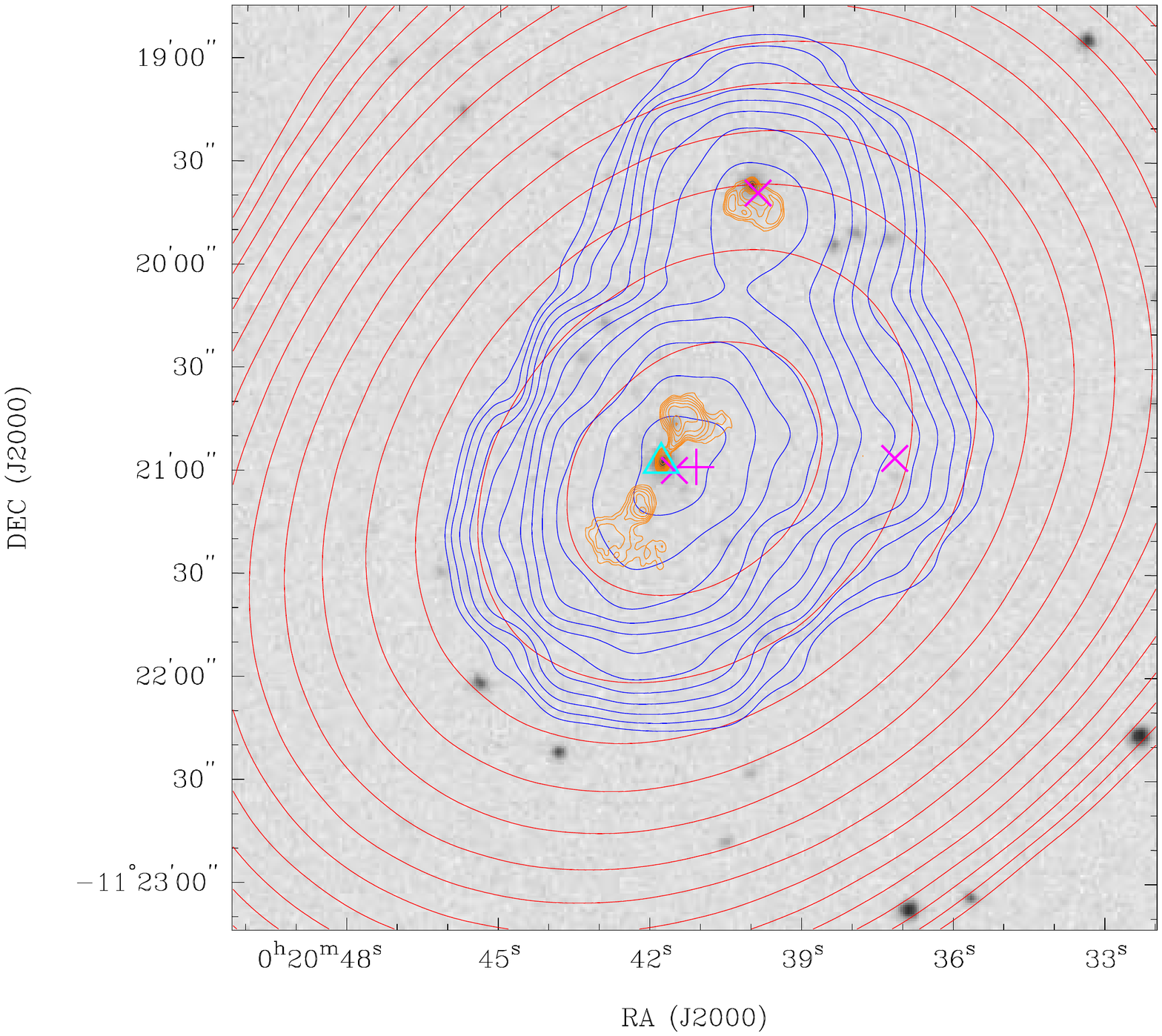}
\includegraphics[scale=0.4, angle=270, trim=0cm 4cm 0cm 4cm]{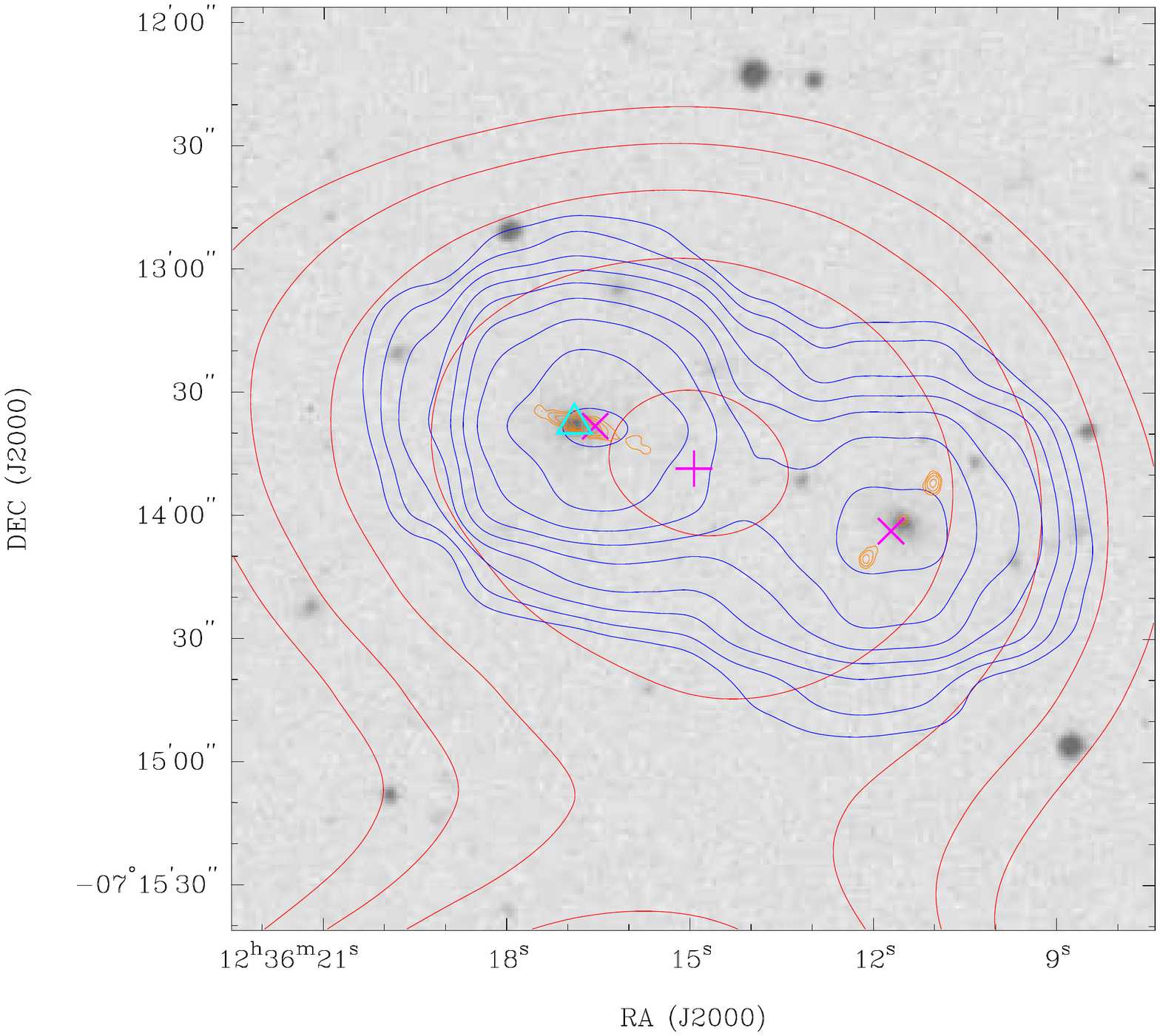}
\caption{Example overlays including higher resolution radio data from VLASS to help determine whether overlapping NVSS components are physically related or not. Radio contours from VLASS (2--4~GHz; orange) are plotted, with the lowest contour at the 5$\sigma$ level and the number of $\sigma$ increasing by a factor of $\sqrt{2}$ with each subsequent contour. The other contours and symbols are as described in Fig.~\ref{fig:example_overlays}. Left: the southern NVSS component is detected as an FRI in VLASS. The northern NVSS component, also detected in VLASS, is physically unrelated. Right: the western NVSS component is detected as a double radio galaxy in VLASS and is physically unrelated with the eastern NVSS component.}
\label{fig:example_vlass}
\end{center}
\end{figure*}

Where necessary, we use higher resolution radio data from FIRST, VLASS and TGSS to help identify by eye the correct NVSS/SUMSS components. The VLASS Quick Look images can be particularly useful to determine whether overlapping NVSS sources are physically related or not, as shown in Fig.~\ref{fig:example_vlass}. In the final GLEAM-6dFGS catalogue, a total of 372 sources (23 per cent) are associated with more than one NVSS/SUMSS component. The source (GLEAM~J225615-361754) with the highest number of components is a giant radio galaxy with ten NVSS components, an angular size of $\approx 14$~arcmin and a linear size of $\approx 1.4$~Mpc.

We identify 13 sources that are associated with more than one GLEAM component; the highest number of GLEAM components is four. As expected, the number of multi-component sources in GLEAM is much smaller due to the larger GLEAM beam size of $\approx 2$~arcmin. We sum the integrated flux densities of the individual GLEAM components to obtain the total, integrated flux density at 200~MHz and at each of the 20 sub-band frequencies. When calculating the errors on the total flux densities, we assume that the component flux density errors are correlated. We provide the name(s) of any associated GLEAM component(s) in the GLEAM-6dFGS catalogue.

\subsubsection{Correcting the GLEAM flux densities for confusion}\label{Correcting the GLEAM flux densities for confusion}

\begin{figure}
\begin{center}
\includegraphics[scale=0.35, angle=270, trim=0cm 2cm 0cm 2cm]{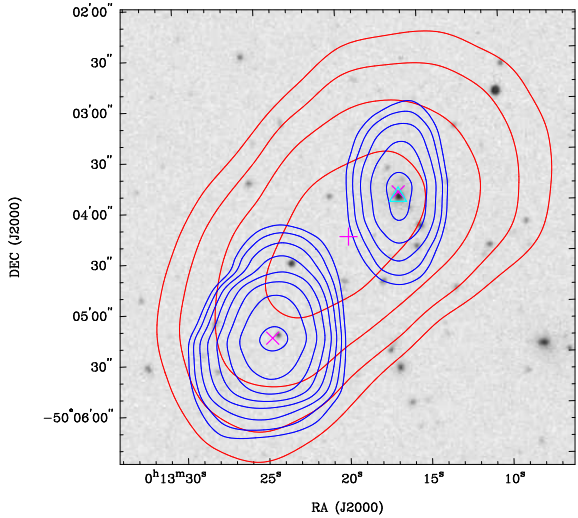}
\caption{Example overlay where two distinct galaxies detected in SUMSS are confused in GLEAM. The contours and symbols are as described in Fig.~\ref{fig:example_overlays}.}
\label{fig:example_confusion}
\end{center}
\end{figure}

From the visual inspection, we find instances where the MWA beam has blended unrelated NVSS/SUMSS sources together. We identify 254 sources whose GLEAM flux densities are likely to be significantly overestimated due to confusion and that require re-fitting. An example of such a source is shown in Fig.~\ref{fig:example_confusion}.

For each of these sources, we use the priorised fitting mode of the \textsc{aegean} source finder \citep{hancock2012,hancock2018} to deblend the GLEAM emission from the 6dFGS galaxy. We extract flux densities from the GLEAM wide-band and sub-band images at the positions of all catalogued NVSS/SUMSS components within 10~arcmin from the 6dFGS galaxy, simultaneously. The flux densities of the components are measured using Gaussian fitting. Only the peak flux densities of the components are allowed to vary in the Gaussian fitting; the positions of the components are fixed and their shapes are set to the synthesised beam of the GLEAM image. This is because the Gaussian fitting can be poorly constrained when allowing the positions and shapes of the components to vary. In any case, given the higher resolution of NVSS and SUMSS, the emission in GLEAM can generally be very well modelled with point sources at the positions of the NVSS and SUMSS components.

The deblended flux density of the NVSS/SUMSS component associated with the 6dFGS galaxy is then taken as our new estimate of the GLEAM flux density at each frequency. If more than one NVSS/SUMSS component is associated with the 6dFGS galaxy, we sum the component flux densities to obtain the total, integrated flux density. After applying this procedure, we obtain deblended flux densities at 20 frequencies between 76 and 227~MHz from the sub-band images. We also obtain deblended flux densities at 200 and 216~MHz from the wide-band images in the shallow and deep regions respectively. The intra-band spectral index is re-measured by fitting a power law to the 20 sub-band flux densities. In the deep region, a best estimate of the 200~MHz deblended flux density is derived using the same method as in the original GLEAM SGP catalogue (see Section~\ref{GLEAM Exgal/SGP catalogue and images}).

For about one third of the sources that are refitted, the deblended 200~MHz flux density falls below the 200-MHz flux density limit of the GLEAM-6dFGS sample (100~mJy in the shallow region and 55~mJy in the deep region). After removing these sources, the final number of sources in the GLEAM-6dFGS sample is 1,590. In the GLEAM-6dFGS catalogue, we use the `refitted flag' to indicate which sources have refitted GLEAM measurements. We also flag sources for which the GLEAM flux densities, and in some cases also the NVSS/SUMSS flux densities, are judged to be unreliable due to severe confusion (`confusion flag') based on the visual inspection. A total of 24 sources are flagged in this way; their flux densities are likely to be overestimated.

\subsection{Cross-matching with AllWISE}\label{Cross-matching with AllWISE}

We use the `CDS Upload X-Match' facility of \textsc{topcat} to cross-match the GLEAM-6dFGS sample with the AllWISE catalogue. At least one match within 5~arcsec of the 6dFGS optical position is obtained for 1,575 of the 1,590 GLEAM-6dFGS sources. In the final GLEAM-6dFGS catalogue, we record the name, position and mid-infrared magnitudes (and errors) of the closest AllWISE match within 5~arcsec of the 6dFGS optical position.

\subsection{Final catalogue description}\label{Final catalogue description}

In Table~\ref{tab:columns}, we describe the 87 columns of the GLEAM-6dFGS source catalogue. The electronic version of the catalogue is available from VizieR.

We use the GLEAM and NVSS/SUMSS data to calculate two-point spectral indices
\begin{equation}
\alpha_{\mathrm{high}} = \frac{\ln[ S(\nu_{1}) / S(\nu_{2}) ]} {\beta} \, ,
\end{equation}
where $\nu_{1} = 200$~MHz,
\begin{equation}
\nu_{2} = 
\begin{cases}
  843~\mathrm{MHz} & \text{if Dec $< -39.5\deg$} \\
  1400~\mathrm{MHz} & \text{otherwise},
\end{cases}
\end{equation}
$S(\nu_{1})$ is the integrated flux density at $\nu_{1}$ from GLEAM, $S(\nu_{2})$ is the integrated flux density at $\nu_{2}$ from NVSS/SUMSS and $\beta = \ln(\nu_{1}/\nu_{2})$. The error on $\alpha_{\mathrm{high}}$ is given by
\begin{equation}
\Delta \alpha_{\mathrm{high}} = \frac{ \sqrt{ \left[ \frac{ \Delta S(\nu_{1}) }{ S(\nu_{1}) } \right]^{2} + \left[ \frac{ \Delta S(\nu_{2}) }{ S(\nu_{2}) } \right]^{2} } } {\beta} \, .
\end{equation}
The 200~MHz flux density errors quoted in the GLEAM Exgal and SGP catalogues, $\sigma_{\mathrm{fit}}(\nu_{1})$, only account for the image noise. We therefore set
\begin{equation}
\Delta S(\nu_{1}) = \sqrt{\sigma_{\mathrm{fit}}(\nu_{1})^{2} + [\epsilon S(\nu_{1})]^{2}} \, ,
\end{equation}
where $\epsilon$ is the GLEAM external flux scale error. The value of $\epsilon$ is Dec-dependent but at most Decs $\epsilon = 0.08$ (see \citealt{hurleywalker2017} and \citealt{franzen2021} for details).

We calculate rest-frame radio luminosities at 200~MHz using a $k$-correction based on the spectral index, $\alpha$, and assuming the radio emission is synchrotron emission characterised by a power law. The radio luminosity of a source with flux density $S_{\nu}$ at redshift $z$ and luminosity distance $d_{\mathrm{L}}$ is therefore given by 
\begin{equation}
L_{\nu} = \frac{4 \pi d_{\mathrm{L}}^2 S_{\nu}} { (1+z)^{\alpha+1}} \, .
\end{equation}
In the radio $k$-correction, we use the spectral index calculated from the 20 GLEAM sub-band flux densities between 76 and 227~MHz, $\alpha_{\mathrm{low}}$, if $\Delta \alpha_{\mathrm{low}} \leq 0.2$. This condition is met for 84 per cent of the sources in the catalogue. For the remaining sources with $\Delta \alpha_{\mathrm{low}} > 0.2$ or no measurement of $\alpha_{\mathrm{low}}$, we use $\alpha_{\mathrm{high}}$ in the radio $k$-correction.

We calculate absolute $K$-band magnitudes using $K$-band $k$-corrections derived by \cite{glazebrook1995} from the evolutionary synthesis models of \cite{bruzual1993} assuming a delta-function burst of star formation at age 5~Gyr. The absolute $K$-band magnitude is given by
\begin{equation}
M_{\mathrm{K,\,abs}} = M_{\mathrm{K}} - 5 \log_{10} \left(\frac{d_{\mathrm{L}}}{10~\mathrm{pc}}\right) - K(z) \, ,
\end{equation}
where $M_{\mathrm{K}}$ is the apparent $K$-band magnitude and $K(z)$ is the $K$-band $k$-correction given by equation 6 of \citeauthor{glazebrook1995}

\cite{white2020b,white2020a} compiled a complete sample of the `brightest' radio sources ($S_{151\,\mathrm{MHz}} > 4$~Jy) at $\mathrm{Dec} < 30\deg$ from the GLEAM Exgal catalogue, the majority of which are AGN with powerful radio jets. The G4Jy sample consists of 1,863 sources and is over 10 times larger than the revised Third Cambridge Catalogue of Radio Sources \citep[3CRR][]{laing1983}. A total of 47 G4Jy sources are in common with the GLEAM-6dFGS sample. For reference, the final column of the catalogue (`G4Jy\_name') gives the name of the G4Jy source. We note that not all G4Jy sources with 6dFGS hosts are expected to be included in the GLEAM-6dFGS catalogue due to the maximum allowed separation of 50~arcsec between the GLEAM and 6dFGS positions, and the source filtering applied in Section~\ref{Source filtering}.

\setlength{\skip\footins}{2pt}

\begin{landscape}
\begin{table}
\caption{Column numbers, names, units, descriptions and first row entries for the 87 columns in the GLEAM-6dFGS source catalogue. All reported magnitudes are in the Vega system.}
\centering
\begin{tabular}{@{}ccccc@{}}
\hline
Number & Name & Unit & Description & First row entry \\
\hline
1 & 6dFGS\_target\_name & -- & Target name of the object from the 6dFGS data base & g0000141-251113 \\
2 & 6dFGS\_RAJ2000 & $\deg$ & RA of the object from the 2MASS XSC (J2000) & $0.05883$ \\
3 & 6dFGS\_DEJ2000 & $\deg$ & Dec of the object from the 2MASS XSC (J2000) & $-25.18692$ \\
4 & z & -- & 6dFGS measured redshift & 0.0852 \\
5 & Q & -- & 6dFGS redshift quality flag as described in \cite{jones2009} & 4 \\
6 & K & mag & Total apparent \textit{K}-band magnitude & $11.89$ \\
7 & K\_abs & mag & Total absolute \textit{K}-band magnitude & $-25.85$ \\
8 & Spectrum\_class & -- & Classification of the spectrum as defined in Table~\ref{tab:spectral_classification} & Aa \\
9 & AllWISE\_name & -- & Name of the host galaxy in AllWISE & J000014.07-251112.6 \\
10 & AllWISE\_RAJ2000 & $\deg$ & RA of the host galaxy (J2000) & $0.05865$ \\
11 & AllWISE\_DEJ2000 & $\deg$ & Dec of the host galaxy (J2000) & $-25.18684$ \\
12 & W1mag & mag & W1 magnitude of the host galaxy & $12.849$ \\
13 & err\_W1mag & mag & Error on W1 magnitude of the host galaxy & $0.025$ \\
14 & W2mag & mag & W2 magnitude of the host galaxy & $12.760$ \\
15 & err\_W2mag & mag & Error on W2 magnitude of the host galaxy & $0.028$ \\
16 & W3mag & mag & W3 magnitude of the host galaxy & $11.780$ \\
17 & err\_W3mag & mag & Error on W3 magnitude of the host galaxy & $0.261$ \\
18 & W4mag & mag & W4 magnitude of the host galaxy & $8.512$ \\
19 & err\_W4mag & mag & Error on W4 magnitude of the host galaxy & - \\
20 & GLEAM\_name & hh:mm:ss+dd:mm:ss & Name of the GLEAM component & GLEAM~J000012-251112 \\
21 & GLEAM\_RAJ2000 & $\deg$ & RA of the GLEAM component (J2000) & $0.05394$ \\
22 & GLEAM\_DEJ2000 & $\deg$ & Dec of the GLEAM component (J2000) & $-25.18671$ \\
23 & Offset & arcsec & Offset between GLEAM and 6dFGS positions & $16.0$ \\
24 & region & -- & Region of the GLEAM survey (deep or shallow) & deep \\
25 & ncmp\_GLEAM & -- & Number of GLEAM components & 1 \\
26 & GLEAM\_associated\_name & -- & Name(s) of any associated GLEAM component(s) & - \\
27 & GLEAM\_err\_abs\_flux\_pct & -- & Percentage error in absolute flux scale & $8.0$ \\
28 & GLEAM\_err\_fit\_flux\_pct & -- & Percentage error in internal flux scale & $2.0$ \\
29 & S\_200 & mJy & Integrated flux density at 200~MHz from GLEAM & $115.8$ \\
30 & dS\_200 & mJy & Error on integrated flux density at 200~MHz from GLEAM & $1.6$ \\
31 & S\_076 & mJy & Integrated flux density at 76~MHz from GLEAM & $246.1$ \\
32 & dS\_076 & mJy & Error on integrated flux density at 76~MHz from GLEAM & $55.7$ \\
33 & S\_084 & mJy & Integrated flux density at 84~MHz from GLEAM & $232.2$ \\
34 & dS\_084 & mJy & Error on integrated flux density at 84~MHz from GLEAM & $42.7$ \\
35 & S\_092 & mJy & Integrated flux density at 92~MHz from GLEAM & $246.7$ \\
36 & dS\_092 & mJy & Error on integrated flux density at 92~MHz from GLEAM & $37.5$ \\
37 & S\_099 & mJy & Integrated flux density at 99~MHz from GLEAM & $243.7$ \\
38 & dS\_099 & mJy & Error on integrated flux density at 99~MHz from GLEAM & $34.6$ \\
39 & S\_107 & mJy & Integrated flux density at 107~MHz from GLEAM & $213.8$ \\
40 & dS\_107 & mJy & Error on integrated flux density at 107~MHz from GLEAM & $21.5$ \\
\label{tab:columns}
\end{tabular}
\end{table}
\end{landscape}

\setcounter{table}{3}

\begin{landscape}
\begin{table}[h!]
\caption{Continued}
\centering
\begin{tabular}{@{}ccccc@{}}
\hline
Number & Name & Unit & Description & First row entry \\
\hline
41 & S\_115 & mJy & Integrated flux density at 115~MHz from GLEAM & $191.4$ \\
42 & dS\_115 & mJy & Error on integrated flux density at 115~MHz from GLEAM & $16.3$ \\
43 & S\_122 & mJy & Integrated flux density at 122~MHz from GLEAM & $186.7$ \\
44 & dS\_122 & mJy & Error on integrated flux density at 122~MHz from GLEAM & $14.6$ \\
45 & S\_130 & mJy & Integrated flux density at 130~MHz from GLEAM & $149.0$ \\
46 & dS\_130 & mJy & Error on integrated flux density at 130~MHz from GLEAM & $12.4$ \\
47 & S\_143 & mJy & Integrated flux density at 143~MHz from GLEAM & $136.5$ \\
48 & dS\_143 & mJy & Error on integrated flux density at 143~MHz from GLEAM & $10.3$ \\
49 & S\_151 & mJy & Integrated flux density at 151~MHz from GLEAM & $144.0$ \\
50 & dS\_151 & mJy & Error on integrated flux density at 151~MHz from GLEAM & $8.6$ \\
51 & S\_158 & mJy & Integrated flux density at 158~MHz from GLEAM & $142.1$ \\
52 & dS\_158 & mJy & Error on integrated flux density at 158~MHz from GLEAM & $8.0$ \\
53 & S\_166 & mJy & Integrated flux density at 166~MHz from GLEAM & $137.0$ \\
54 & dS\_166 & mJy & Error on integrated flux density at 166~MHz from GLEAM & $7.3$ \\
55 & S\_174 & mJy & Integrated flux density at 174~MHz from GLEAM & $141.1$ \\
56 & dS\_174 & mJy & Error on integrated flux density at 174~MHz from GLEAM & $7.2$ \\
57 & S\_181 & mJy & Integrated flux density at 181~MHz from GLEAM & $126.7$ \\
58 & dS\_181 & mJy & Error on integrated flux density at 181~MHz from GLEAM & $5.9$ \\
59 & S\_189 & mJy & Integrated flux density at 189~MHz from GLEAM & $126.2$ \\
60 & dS\_189 & mJy & Error on integrated flux density at 189~MHz from GLEAM & $6.0$ \\
61 & S\_197 & mJy & Integrated flux density at 197~MHz from GLEAM & $121.8$ \\
62 & dS\_197 & mJy & Error on integrated flux density at 197~MHz from GLEAM & $5.6$ \\
63 & S\_204 & mJy & Integrated flux density at 204~MHz from GLEAM & $113.0$ \\
64 & dS\_204 & mJy & Error on integrated flux density at 204~MHz from GLEAM & $5.0$ \\
65 & S\_212 & mJy & Integrated flux density at 212~MHz from GLEAM & $115.0$ \\
66 & dS\_212 & mJy & Error on integrated flux density at 212~MHz from GLEAM & $4.7$ \\
67 & S\_220 & mJy & Integrated flux density at 220~MHz from GLEAM & $100.8$ \\
68 & dS\_220 & mJy & Error on integrated flux density at 220~MHz from GLEAM & $4.3$ \\
69 & S\_227 & mJy & Integrated flux density at 227~MHz from GLEAM & $101.9$ \\
70 & dS\_227 & mJy & Error on integrated flux density at 227~MHz from GLEAM & $4.3$ \\
71 & alpha\_low & -- & Spectral index between 76 and 227~MHz from GLEAM & $-0.86$ \\
72 & dalpha\_low & -- & Error on spectral index between 76 and 227~MHz from GLEAM & $0.05$ \\
73 & reduced\_chi2\_alpha\_low & -- & Reduced $\chi^{2}$ value for GLEAM spectral index fit\tablefootnote{The reduced $\chi^2$ value from the power-law fit can be used to assess the quality of the fitted spectral index: for 18 degrees of freedom, $P$(reduced $\chi^2 > 1.93$)~$<1$\% and $P$(reduced $\chi^2 > 2.35$)~$<0.1$\%.} & $0.72$ \\
74 & refitted\_flag & -- & Flag for sources with refitted GLEAM measurements & 0 \\
75 & confusion\_flag & -- & Flag for sources severely affected by confusion in GLEAM & 0 \\
\label{tab:columns}
\end{tabular}
\end{table}
\end{landscape}

\setcounter{table}{3}

\begin{landscape}
\begin{table}[h!]
\caption{Continued}
\centering
\begin{tabular}{@{}ccccc@{}}
\hline
Number & Name & Unit & Description & First row entry \\
\hline
76 & P\_200 & W~$\mathrm{Hz}^{-1}$ & Logarithm of the radio luminosity at 200~MHz & $24.31$ \\
77 & nmp\_NVSSorSUMSS & -- & Number of NVSS/SUMSS components & 1 \\
78 & centroid\_RAJ2000 & $\deg$ & RA of the centroid position from NVSS/SUMSS (J2000) & $0.05754$ \\
79 & err\_centroid\_RAJ2000 & arcsec & Error on the RA of the centroid position from NVSS/SUMSS & $0.81$ \\
80 & centroid\_DEJ2000 & $\deg$ & Dec of the centroid position from NVSS/SUMSS (J2000) & $-25.18719$ \\
81 & err\_centroid\_DEJ2000 & arcsec & Error on the Dec of the centroid position from NVSS/SUMSS & $0.90$ \\
82 & S\_NVSSorSUMSS & mJy & Integrated flux density at 1400/843~MHz from NVSS/SUMSS & $28.4$ \\
83 & dS\_NVSSorSUMSS & mJy & Error on integrated flux density at 1400/843~MHz from NVSS/SUMSS & $1.5$ \\
84 & alpha\_high & -- & Spectral index between 200 and 1400/843~MHz & $-0.72$ \\
85 & dalpha\_high & -- & Error on spectral index between 200 and 1400/843~MHz & $0.05$ \\
86 & Freq & MHz & Indicates whether NVSS or SUMSS is used & $1400$ \\
87 & G4Jy\_name & -- & Name of the source from the G4Jy sample & - \\
\hline
\label{tab:columns}
\end{tabular}
\end{table}
\end{landscape}

\section{Radio properties of the GLEAM-6\lowercase{d}FGS sample} \label{Radio properties of the GLEAM-6dFGS sample}
 
Table~\ref{tab:spectral_classification} shows the distribution by type of the spectral classification of the GLEAM-6dFGS sample. Of the 1,590 sources, 1,157 (72.8 per cent) are classified as AGN and 427 (26.9 per cent) as SF galaxies. The vast majority of the AGN have pure absorption-line spectra typical of giant elliptical galaxies. For six sources (0.4 per cent), the optical spectrum is unclassifiable or unknown; the analysis presented in this section does not include the six sources with `unknown' optical spectra.

\begin{table}
\centering
\caption{Spectral classes of the GLEAM-6dFGS objects.}
\label{tab:spectral_classification}
\begin{tabular}{@{} c c c}
\hline
Class & Type of spectrum & Number \\
\hline
 Aa & Pure absorption-line spectrum & 933 (58.7\%) \\ 
 Ae & Strong narrow emission lines & 118 (7.4\%) \\ 
 Aae & Weak narrow emission lines & 106 (6.7\%) \\ 
 SF & HII region-like emission spectrum & 427 (26.9\%) \\ 
 ? & Unknown & 6 (0.4\%) \\ 
 & & \\ 
 Total & & 1,590 \\
\hline
\end{tabular}
\end{table}

\cite{best2012} proposed a fundamental dichotomy between high-excitation radio galaxies (HERGs), fuelled at high rates through radiatively-efficient classic accretion disks, and low-excitation radio galaxies (LERGs), fuelled at significantly lower rates via radiatively-inefficient flows. Observationally, HERGs show strong optical emission lines relative to the stellar continuum while LERGs show weak or no optical emission lines. In their study of the local radio source population at 20~GHz, \cite{sadler2013} made a qualitative separation by associating `Ae' radio galaxies with HERGs, and `Aa' and `Aae' radio galaxies with LERGs. If we make the same separation, 10 per cent of the AGN in the GLEAM-6dFGS sample are classified as HERGs.

\begin{figure}
\includegraphics[scale=0.52, trim=0cm 0cm 0cm 0cm]{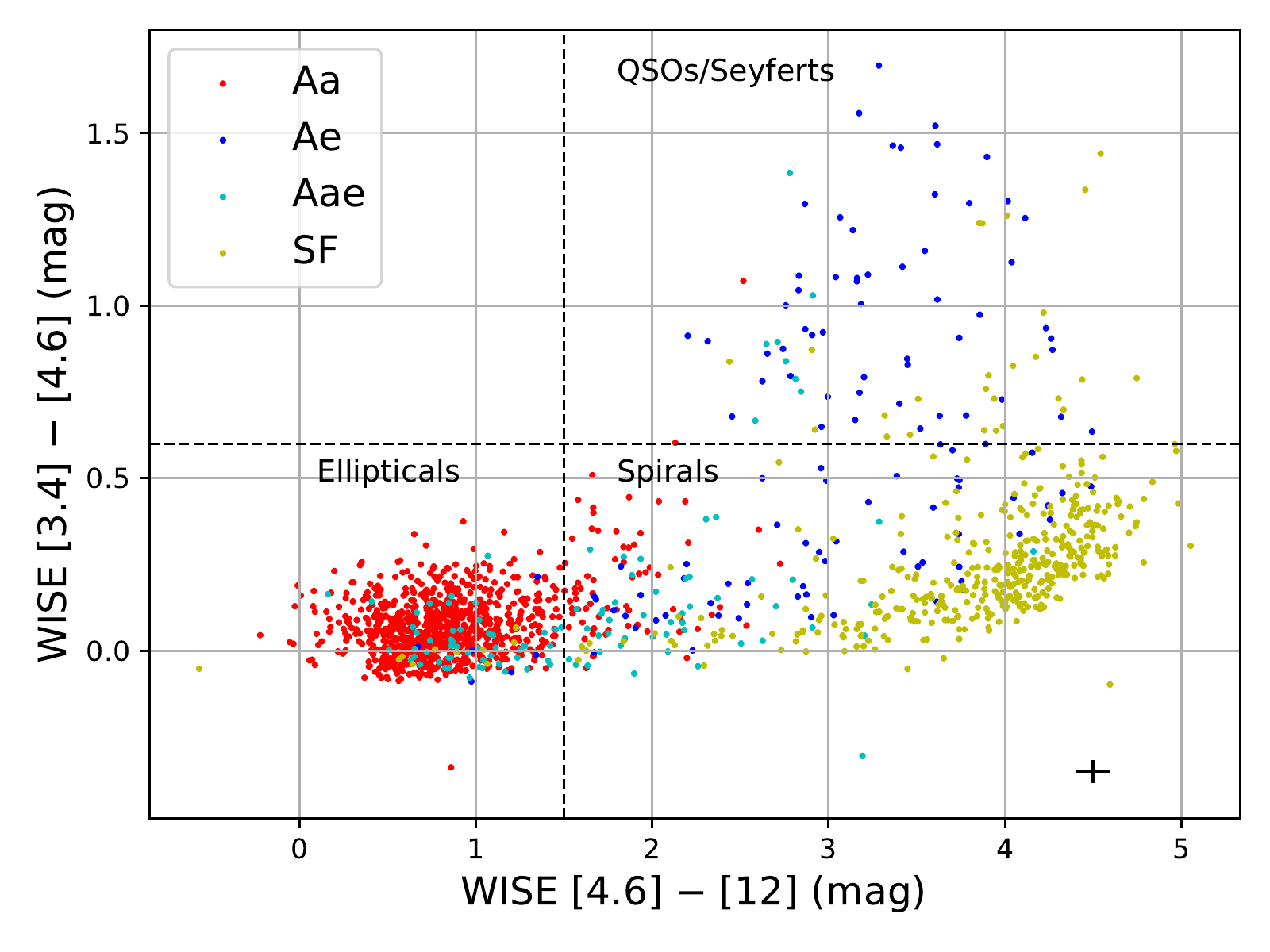}
\caption{WISE colour-colour plot for the GLEAM-6dFGS galaxies, which are colour-coded based on the type of optical spectrum. The WISE magnitudes are in the Vega system. The horizontal and vertical dashed lines divide the different galaxy populations, as discussed in the text. Individual error bars are not plotted for clarity but the median error bar size for the sample is shown at the bottom right.}
\label{fig:WISE_colour_colour}
\end{figure}

In Fig.~\ref{fig:WISE_colour_colour} we report a WISE colour-colour plot \citep{wright2010} in W1 ($3.4~\mu$m), W2 ($4.6~\mu$m) and W3 ($12~\mu$m). The median errors on W1--W2 and W2--W3 are 0.033 and 0.10 respectively. The dashed horizontal line at a W1--W2 colour of $+0.6$~mag shows the dividing line between normal galaxies (LERGs) and radiatively-efficient AGN (HERGs), and the vertical dashed line at a W2--W3 colour of $+1.5$~mag shows the dividing line between elliptical and spiral galaxies, both adopted by \citeauthor{wright2010}

The WISE galaxy classes are broadly consistent with the visual classification of the optical spectra: only 2.4 per cent of the SF galaxies lie in the `WISE elliptical' region. However, 53 per cent of the Ae galaxies lie in the `WISE LERG' region. In a study of the WISE properties of a large sample of radio-identified galaxies and QSOs, \cite{ching2017} noted that the Ae/HERG objects are a strongly heterogeneous class spanning a wide area in the WISE two-colour diagram. They concluded that the physical process dominating the optical/MIR light is not the same for all members of this class, and that the position of individual HERGs in the WISE two-colour plot is likely to depend on the star formation rate within the host galaxy, the brightness of the optical AGN relative to the host galaxy, and the amount of dust obscuration present within the galaxy. Our results for the 6dFGS radio-detected galaxies are consistent with this picture.  

\subsection{Flux density distribution}\label{Flux density distribution}

\begin{figure}
\begin{center}
\includegraphics[scale=0.55, trim=0cm 0cm 0cm 0cm]{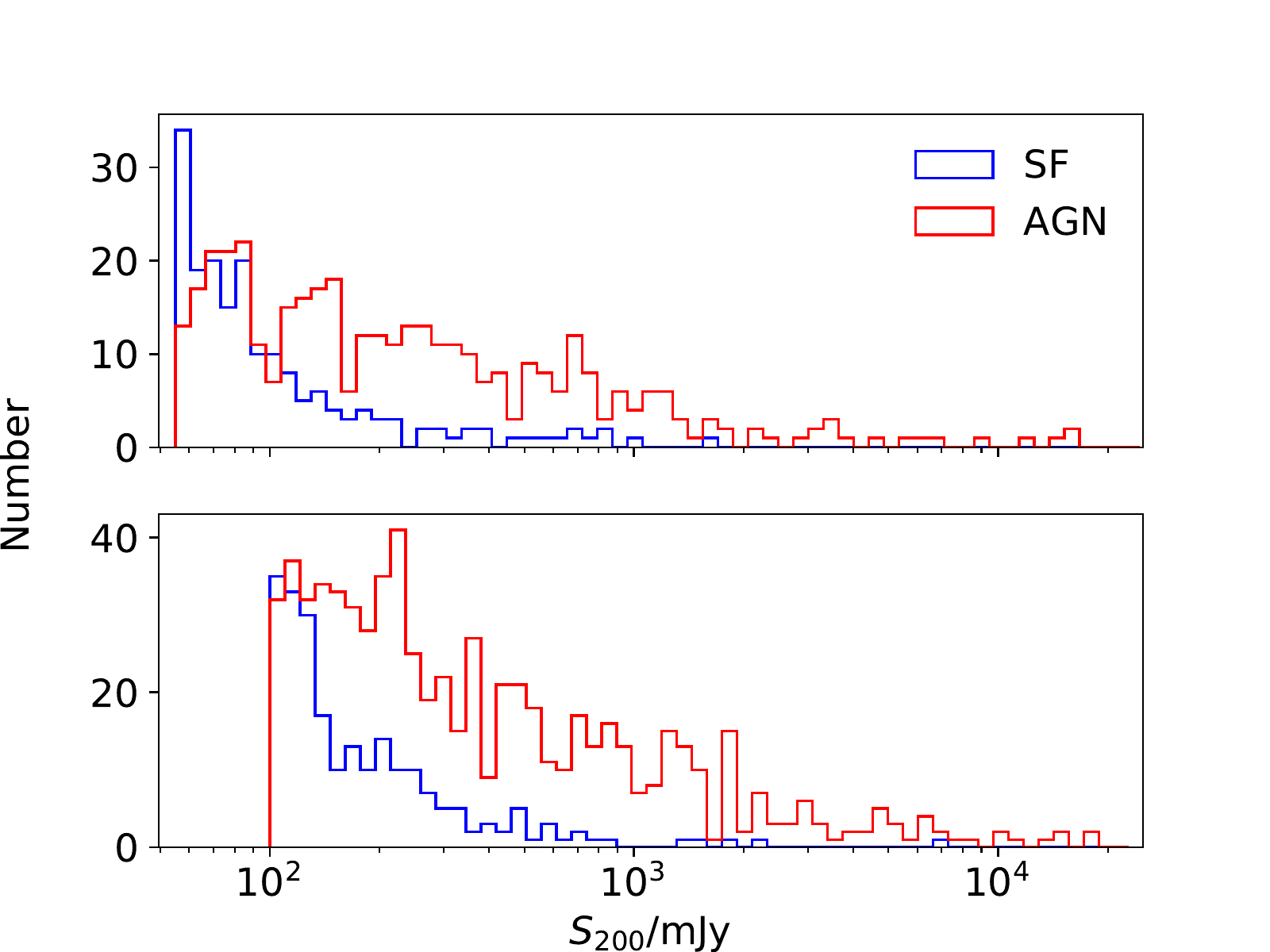}
\caption{Histogram of $S_{200\,\mathrm{MHz}}$ for the AGN (red) and SF galaxies (blue) in the deep (top) and shallow (bottom) regions of the GLEAM-6dFGS sample. Flux density cuts of 55 and 100~mJy are applied in the deep and shallow regions respectively.}
\label{fig:flux_dist}
\end{center}
\end{figure}

Fig.~\ref{fig:flux_dist} shows the 200~MHz flux density distributions of the AGN and SF galaxies in the deep and shallow regions. The number of AGN and SF galaxies is roughly equal at 100~mJy, the flux density limit in the shallow region. The sample is dominated by SF galaxies close to the flux density limit in the deep region (55~mJy).

\subsection{Redshift and luminosity distributions}\label{Redshift and luminosity distributions}

\begin{figure}
\includegraphics[scale=0.42, trim=0cm 0cm 0cm 0cm]{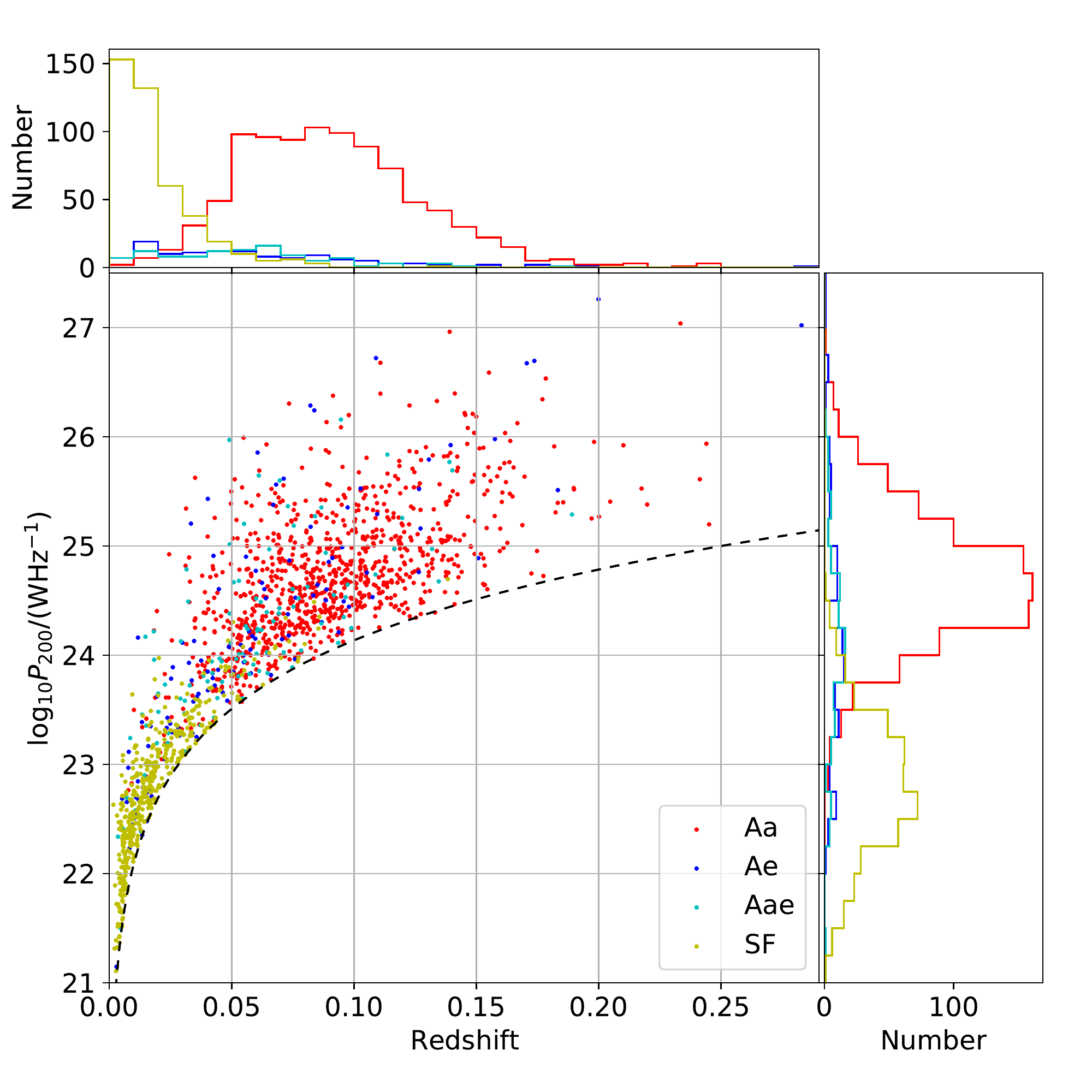}
\caption{200~MHz radio luminosity as a function of redshift for the AGN and SF galaxies in the GLEAM-6dFGS sample. The dashed line corresponds to the 200~MHz flux density limit of 55~mJy in the deep region.}
\label{fig:power_vs_z}
\end{figure}

In Fig.~\ref{fig:power_vs_z}, we present the 200~MHz radio luminosity and redshift distributions of the AGN and SF galaxies. The AGN have a median redshift of 0.081. In comparison, the median redshift of the AGN in the NVSS-6dFGS sample by \cite{mauch2007}, selected at 1.4~GHz, is 0.073. The NVSS is estimated to be 99 per cent complete at $S_{1.4\,\mathrm{GHz}} = 3.4$~mJy. The equivalent completeness limit of NVSS at 200~MHz for a compact radio source with a spectral index of --0.7 is $\approx 13$~mJy. This is a factor of $\approx 4-8$ lower than the 200-MHz flux density limits of the GLEAM-6dFGS sample. The fact that the median redshift of the AGN in the NVSS-6dFGS sample is similar to that in the GLEAM-6dFGS sample implies that AGN drop out of the GLEAM-6dFGS sample primarily due to the $K$-band magnitude limit. 

The SF galaxies form a more nearby population with a median redshift of 0.015. The median redshift (0.035) of the SF galaxies in the NVSS-6dFGS sample is more than twice as high as that in the GLEAM-6dFGS sample, indicating that it is mainly the radio flux density in GLEAM that limits the maximum distance to which SF galaxies can be detected.

As expected, there is a strong separation between the two source populations according to the radio luminosity, as seen in studies at 1.4~GHz. The AGN have median $P_{200\,\mathrm{MHz}} = 10^{24.61}~\mathrm{W}~\mathrm{Hz}^{-1}$ and the SF galaxies have median $P_{200\,\mathrm{MHz}} = 10^{22.79}~\mathrm{W}~\mathrm{Hz}^{-1}$. Among the AGN, 82 per cent have $P_{200\,\mathrm{MHz}} > 10^{24}~\mathrm{W}~\mathrm{Hz}^{-1}$ while 97 per cent of the SF galaxies have $P_{200\,\mathrm{MHz}} < 10^{24}~\mathrm{W}~\mathrm{Hz}^{-1}$. There is a strong increase in the fraction of sources with optical emission lines with decreasing luminosity: while only 19 per cent of all AGN are classed as `Aae' or `Ae', this fraction rises to 51 per cent among the population of low-luminosity ($P_{200\,\mathrm{MHz}} < 10^{24}~\mathrm{W}~\mathrm{Hz}^{-1}$) AGN overlapping in radio luminosity with the SF galaxies.

\begin{figure*}
\begin{center}
\includegraphics[scale=0.55, trim=0cm 0cm 0cm 0cm]{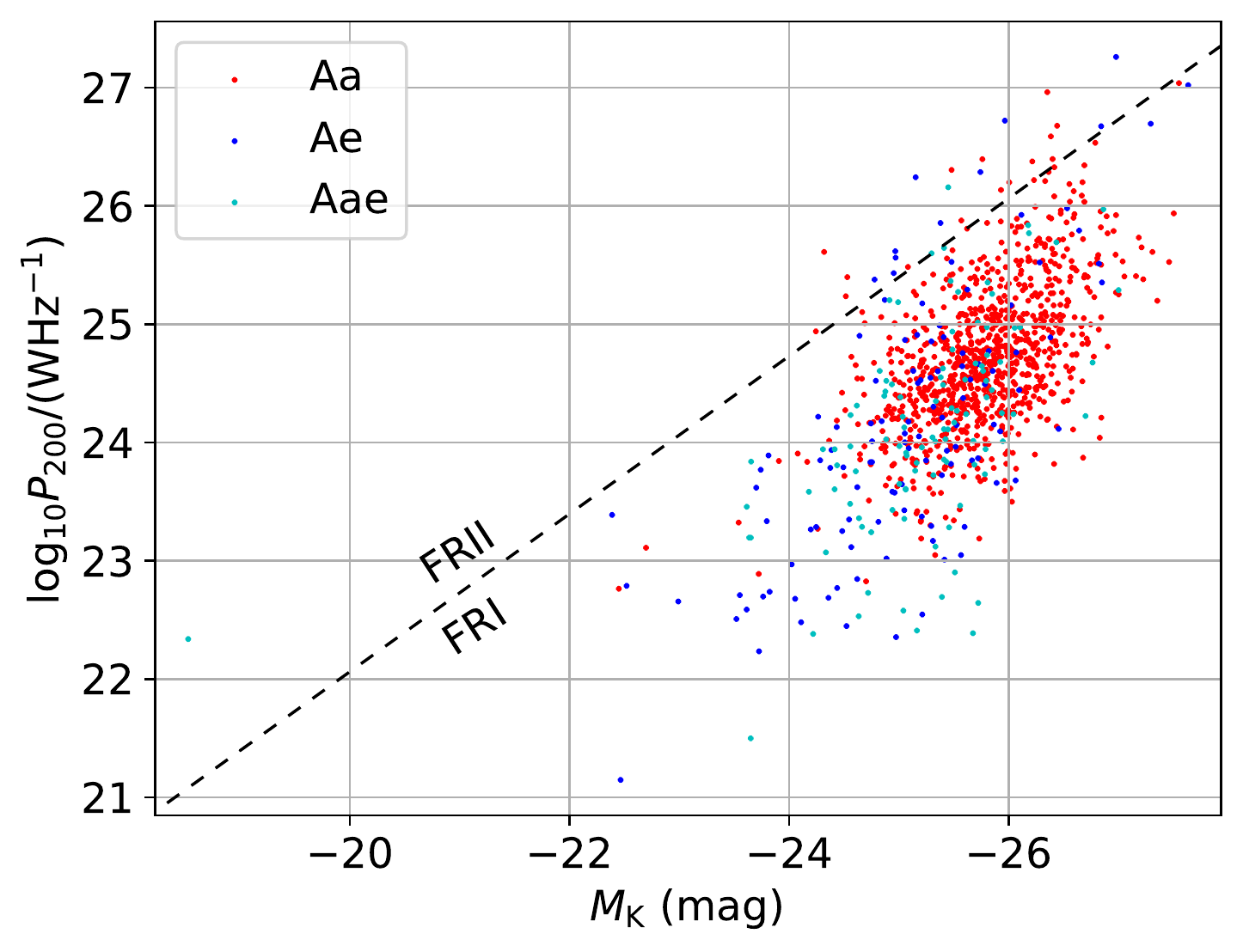}
\includegraphics[scale=0.55, trim=0cm 0cm 0cm 0cm]{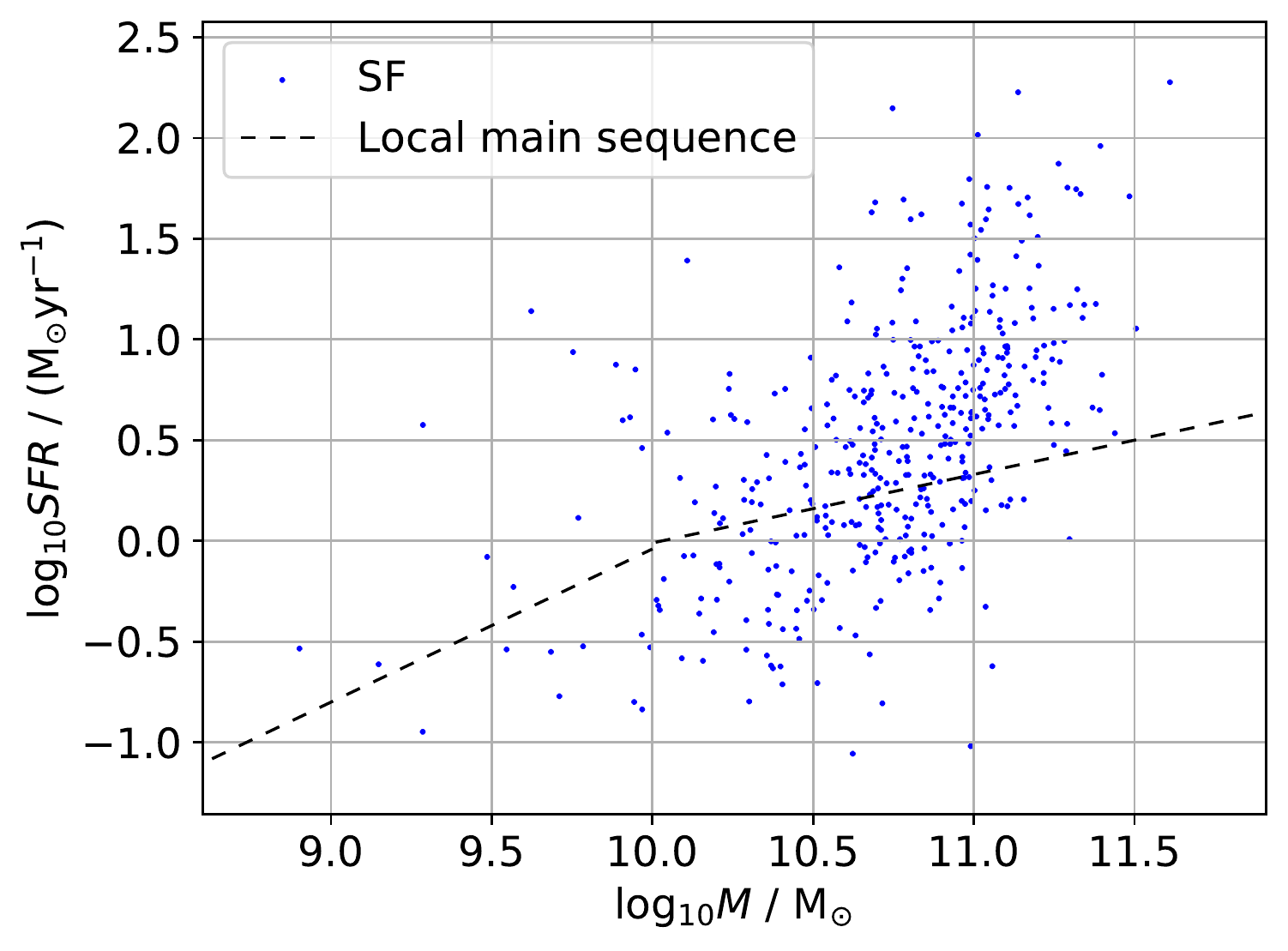}
\caption{Left: distribution in $K$-band absolute magnitude and 200~MHz radio power for the AGN in the GLEAM-6dFGS sample. The dashed line corresponds to the FRI/FRII division at 1.4~GHz from \citet{ledlow1996}, extrapolated to $K$-band and 200~MHz as described in the text, with FRII galaxies lying above the line. Right: distribution in the SFR and stellar mass for the SF galaxies in the GLEAM-6dFGS sample. The stellar mass is derived from the $K$-band absolute magnitude and the SFR from the 200~MHz radio power and stellar mass, as described in the text. The dashed line shows the SF main sequence for galaxies in the local universe by \cite{renzini2015} and \cite{popesso2019}.}
\label{fig:power_vs_KbandABS}
\end{center}
\end{figure*}

The $K$-band absolute magnitude as a function of the 200~MHz radio luminosity for the AGN is shown in the left panel of Fig.~\ref{fig:power_vs_KbandABS}. The dashed line corresponds to the 1.4~GHz FRI/FRII dividing line from \citet{ledlow1996}, shifted to $K$-band and 200~MHz assuming a typical galaxy colour of $R-K = 3.0$~mag and a typical radio spectral index $\alpha_{200}^{1400} = -0.7$. According to this relation, our sample of nearby AGN is overwhelmingly dominated by FRI galaxies: only 29 of the 1157 AGN (2.5 per cent) lie above the dividing line. We note, however, that there have been a number of works showing that this division line is not nearly as clean as \citeauthor{ledlow1996} suggested \citep[see e.g.][]{mingo2019}. Secondly, our sample may contain a significant fraction of compact radio sources which are often referred to as FR0 galaxies.

Since the low-frequency radio emission from SF galaxies is primarily due to the acceleration of electrons via shocks associated with supernovae, the radio luminosity from SF galaxies can be used as a tracer of the star formation rate \citep[SFR; e.g.][]{condon1992}. \cite{smith2021} studied the relationship between 150~MHz luminosity and SFR using LOFAR measurements at 150~MHz for a near-infrared selected sample of $z<1$ galaxies. We derive SFRs for the SF galaxies in our sample using their best-fit relation, which accounts for the stellar mass, $M_{*}$:
\begin{flalign}
\begin{split}
\log_{10} \left( \frac{P_{150\,\mathrm{MHz}}}{\mathrm{W} \mathrm{Hz}^{-1}} \right) &= 0.90 \log_{10} \left( \frac{SFR}{\mathrm{M}_{\odot} \mathrm{yr}^{-1}} \right)\\
&\hphantom{{}=} +0.33 \log_{10} \left( \frac{M_{*}} {10^{10} \mathrm{M}_{\odot}} \right) + 22.22\,.
\end{split}
\end{flalign}
We extrapolate the 200~MHz radio powers measured in this paper to 150~MHz using $\alpha_{\mathrm{low}}$, if $\Delta \alpha_{\mathrm{low}} < 0.2$, otherwise we use $\alpha_{\mathrm{high}}$. We derive $M_{*}$ from $M_{\mathrm{K,\,abs}}$ by combining equations 4 and 5 of \cite{longhetti2009}:
\begin{flalign}
\begin{split}
\log_{10} \frac{M_{*}} {\mathrm{M}_{\odot}} &= -0.4 (M_{\mathrm{K,\,abs}} - M_{\mathrm{K,\,abs,\,Sun}})\\
&\hphantom{{}=} + \log_{10} \frac{ \Upsilon_{\mathrm{K}}} { \mathrm{M}_{\odot} / \mathrm{L}_{\odot} }\,,
\end{split}
\end{flalign}
where $\Upsilon_{\mathrm{K}}$ is the mass-to-light ratio in the $K$-band and $M_{\mathrm{K,\,abs,\,Sun}}$ is the absolute $K$-band magnitude of the Sun. Since disk/spiral galaxies in the local universe typically have $\Upsilon_{\mathrm{K}} \approx 0.6$ \citep{mcgaugh2014} and $M_{\mathrm{K,\,abs,\,Sun}} \approx 3.27$~mag in the Vega system \citep{willmer2018},
\begin{equation}
\log_{10} \frac{M_{*}} {\mathrm{M}_{\odot}} \approx 1.09-0.4 M_{\mathrm{K,\,abs}}\,.
\end{equation}

The SFR as a function of $M_{*}$ is plotted in the right panel of Fig.~\ref{fig:power_vs_KbandABS}. The dashed line shows the main sequence of SF galaxies defined by \cite{renzini2015} at $M_{*} < 10^{10}~\mathrm{M}_{\odot}$ and by \cite{popesso2019} at $M_{*} \geq 10^{10}~\mathrm{M}_{\odot}$. The SF galaxies in the GLEAM-6dFGS sample scatter around and above the high-mass end of the SF main sequence. This is consistent with them lying mainly in the `starburst' and `LIRG/ULIRG' regions of the WISE colour-colour diagram shown in Fig.~\ref{fig:WISE_colour_colour}.

\subsection{Radio spectra}\label{Radio spectra}

\begin{figure}
\includegraphics[scale=0.52, trim=0cm 0cm 0cm 0cm]{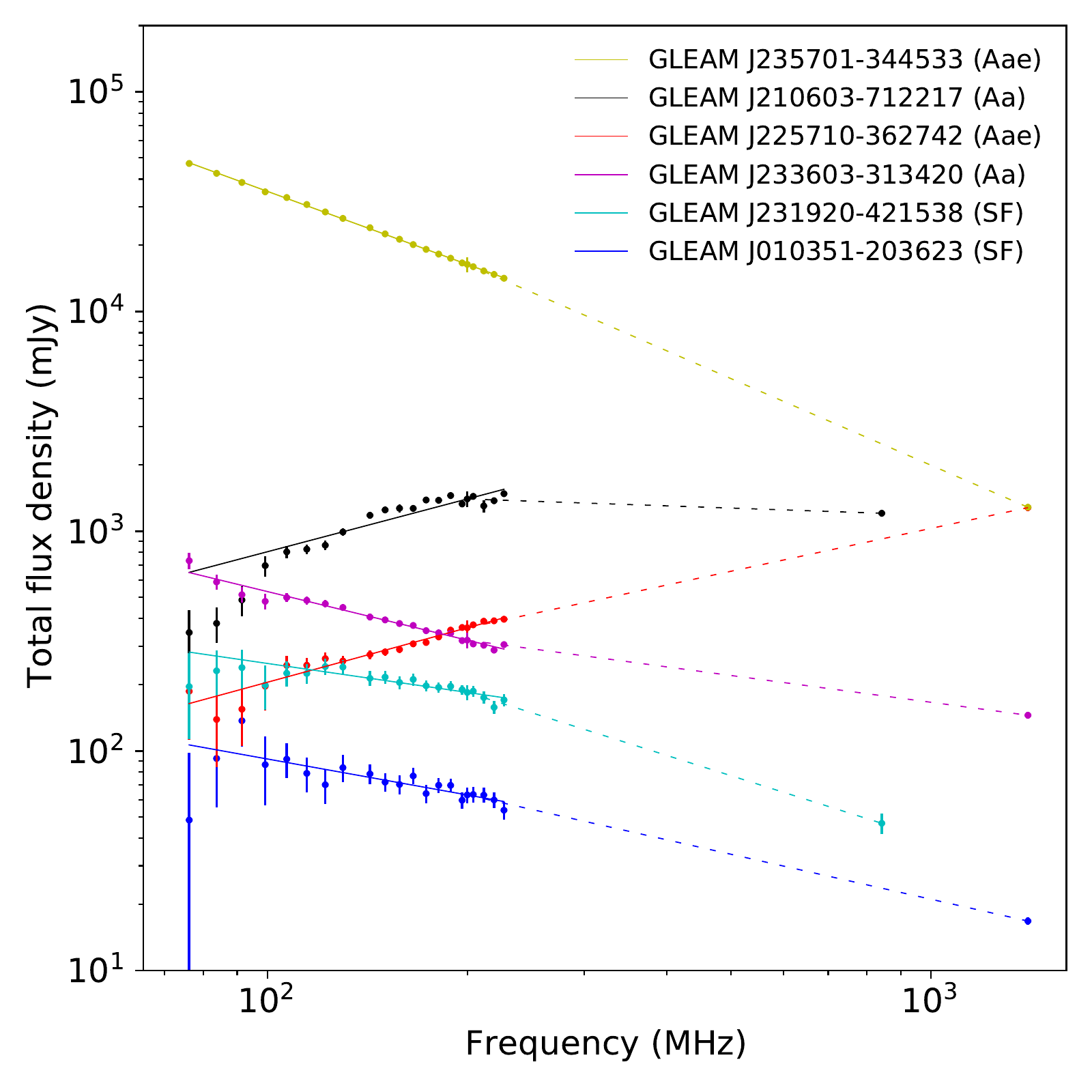}
\caption{Example SEDs using the 20 GLEAM sub-band flux densities between 76 and 227~MHz, the GLEAM wide-band flux density at 200~MHz and the NVSS/SUMSS flux density at 1400/843~MHz. For each source, the solid line shows the power-law fit to the GLEAM sub-band flux densities and the dashed line connects the GLEAM wide-band flux density to the NVSS/SUMSS flux density. Yellow: AGN classed as `Aae' with a steep power-law spectrum across the GLEAM band; black: AGN classed as `Aa' peaking at around 200~MHz; red: AGN classed as `Aae' with a highly inverted spectrum; magenta: AGN classed as `Aa' with a flattening spectrum at high frequency; cyan: SF galaxy with a turnover frequency around 100~MHz; blue: SF galaxy close to the 200-MHz detection limit of the sample with a typical SED.}
\label{fig:example_SED}
\end{figure}

All sources in the GLEAM-6dFGS catalogue have a measurement of $\alpha_{\mathrm{high}}$, while 92 per cent of the sources have a measurement of $\alpha_{\mathrm{low}}$. Fig.~\ref{fig:example_SED} illustrates the wide variety of radio SEDs in the sample.

For $\alpha_{\mathrm{low}}$ to be quoted in the catalogue, the source must have a positive flux density in each of the 20 GLEAM sub-bands, which is not always the case at low signal-to-noise ratio (SNR). Of the 1,590 sources in the GLEAM-6dFGS sample, we investigate the spectral properties of the sub-sample of 1,439 sources that have measured $\alpha_{\mathrm{low}}$ and $\alpha_{\mathrm{high}}$, excluding the 24 sources severely affected by confusion in GLEAM and the six sources with `unknown' optical spectra. We note that this sub-sample is biased against faint sources with flat spectra across the GLEAM frequency range: the negative flux densities mostly occur in the lowest frequency sub-bands which have higher rms noise. It is also biased against diffuse radio sources with ultra-steep radio spectra, as discussed in Section~\ref{Radio completeness}. The median errors on $\alpha_{\mathrm{low}}$ and $\alpha_{\mathrm{high}}$ are 0.08 and 0.06 respectively.

\begin{figure*}
\begin{center}
\includegraphics[scale=1.2, trim=0cm 0cm 0cm 0cm]{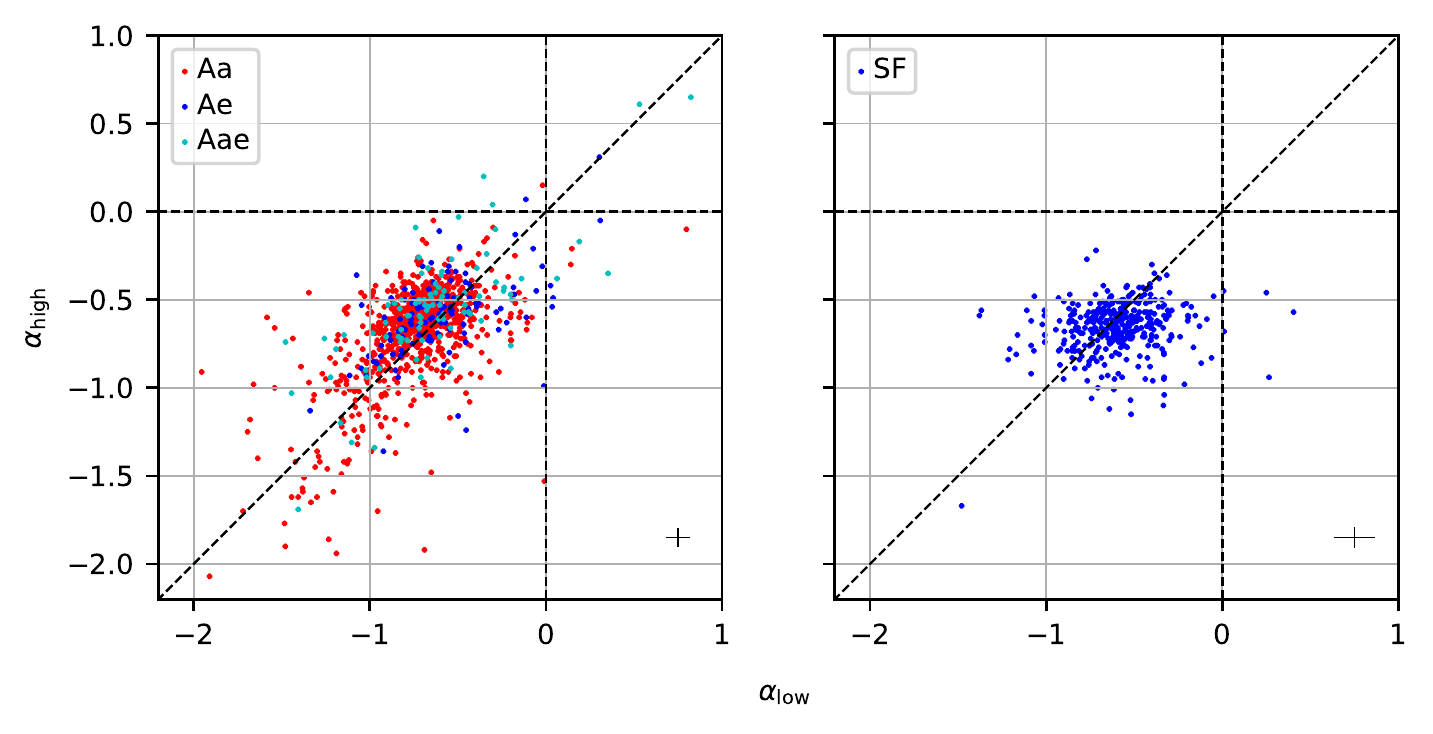}
\caption{Left: radio colour-colour diagram for all AGN in the GLEAM-6dFGS sample with measured $\alpha_{\mathrm{low}}$ and $\alpha_{\mathrm{high}}$; $\alpha_{\mathrm{low}}$ is the spectral index between 76 and 227~MHz, and $\alpha_{\mathrm{high}}$ is the spectral index between 200 and 1400/843~MHz. AGN classed as `Aa' are shown in red, `Ae' in blue and `Aae' in cyan. The dashed lines represent spectral indices of zero and equal values of $\alpha_{\mathrm{low}}$ and $\alpha_{\mathrm{high}}$. Individual error bars are not plotted for clarity but the median error bar size for the sample is shown at the bottom right. Right: radio colour-colour diagram for all SF galaxies in the GLEAM-6dFGS sample with measured $\alpha_{\mathrm{low}}$ and $\alpha_{\mathrm{high}}$.}
\label{fig:colours}
\end{center}
\end{figure*}

The left panel of Fig.~\ref{fig:colours} shows a radio colour-colour plot for all AGN in the sub-sample, which are colour-coded according to their optical spectral classification. The right panel shows a radio colour-colour plot for all SF galaxies in the sub-sample. In Table~\ref{tab:colours}, the median and standard deviation values of $\alpha_{\mathrm{low}}$ and $\alpha_{\mathrm{high}}$ are provided for each class of optical spectrum. Table~\ref{tab:colours} also gives the fractions of sources with $\alpha_{\mathrm{low}} > -0.5$, $\alpha_{\mathrm{high}} > -0.5$, $\alpha_{\mathrm{low}} < -1.2$ and $\alpha_{\mathrm{high}} < -1.2$.

\begin{table*}
\small
\centering
\caption{Spectral index statistics for the AGN and SF galaxies with measured $\alpha_{\mathrm{low}}$ and $\alpha_{\mathrm{high}}$ in the GLEAM-6dFGS sample. $N_{\mathrm{flat}}$ is the number of flat-spectrum sources with $\alpha > -0.5$ and $N_{\mathrm{USS}}$ is the number of ultra-steep-spectrum sources with $\alpha < -1.2$.}
\label{tab:colours}
\begin{tabular}{@{} c c | c c c c | c c c c}
\hline 
& & \multicolumn{4}{c|}{$\alpha_{\mathrm{low}}$}  & \multicolumn{4}{c}{$\alpha_{\mathrm{high}}$} \\
Class & $N_{\mathrm{total}}$ & Median & Std. & $N_{\mathrm{flat}}$ & $N_{\mathrm{USS}}$ & Median & Std. & $N_{\mathrm{flat}}$ & $N_{\mathrm{USS}}$ \\
& & & dev. & & & & dev. & & \\
\hline 
Aa & 869 & $-0.713\pm0.011$ & 0.26 & 107 (12\%) & 39 (4.5\%) & $-0.610\pm0.011$ & 0.27 & 197 (23\%) & 44 (5.1\%) \\ 
Ae & 108 & $-0.612\pm0.036$ & 0.30 & 34 (31\%) & 1 (0.9\%) & $-0.575\pm0.029$ & 0.24 & 34 (31\%) & 2 (1.9\%) \\ 
Aae & 93 & $-0.667\pm0.048$ & 0.37 & 26 (28\%) & 5 (5.4\%) & $-0.530\pm0.042$ & 0.33 & 38 (41\%) & 3 (3.2\%) \\ 
All AGN & 1070 & $-0.704\pm0.011$ & 0.28 & 167 (16\%) & 45 (4.2\%) & $-0.600\pm0.010$ & 0.27 & 269 (25\%) & 49 (4.6\%) \\ 
SF & 369 & $-0.596\pm0.015$ & 0.23 & 111 (30\%) & 5 (1.4\%) & $-0.650\pm0.010$ & 0.15 & 38 (10\%) & 1 (0.3\%) \\
\hline 
\end{tabular}
\end{table*}

\subsubsection{AGN}\label{AGN radio spectra}

For the AGN, the median value of $\alpha_{\mathrm{low}}$ is $-0.704 \pm 0.011$ and the median value of $\alpha_{\mathrm{high}}$ is $-0.600 \pm 0.010$. The slight spectral flattening with frequency is likely due to the radio emission becoming increasingly core dominated at high frequency. In their LoTSS-SDSS sample of nearby galaxies, \cite{sabater2019} measured the 150--1400~MHz spectral indices of 496 radio AGN detected in NVSS and FIRST. For sources with $S_{1400\,\mathrm{MHz}} > 20$~mJy, they obtained a median spectral index of --0.63, which is close to the median value of $\alpha_{\mathrm{high}}$ for the AGN in the GLEAM-6dFGS sample. The comparison is performed over a similar flux density range since $\approx 90$ per cent of the AGN in the GLEAM-6dFGS sample have NVSS/SUMSS flux densities above 20~mJy.

\begin{figure*}
\begin{center}
\includegraphics[scale=0.7, trim=0cm 0cm 0cm 0cm]{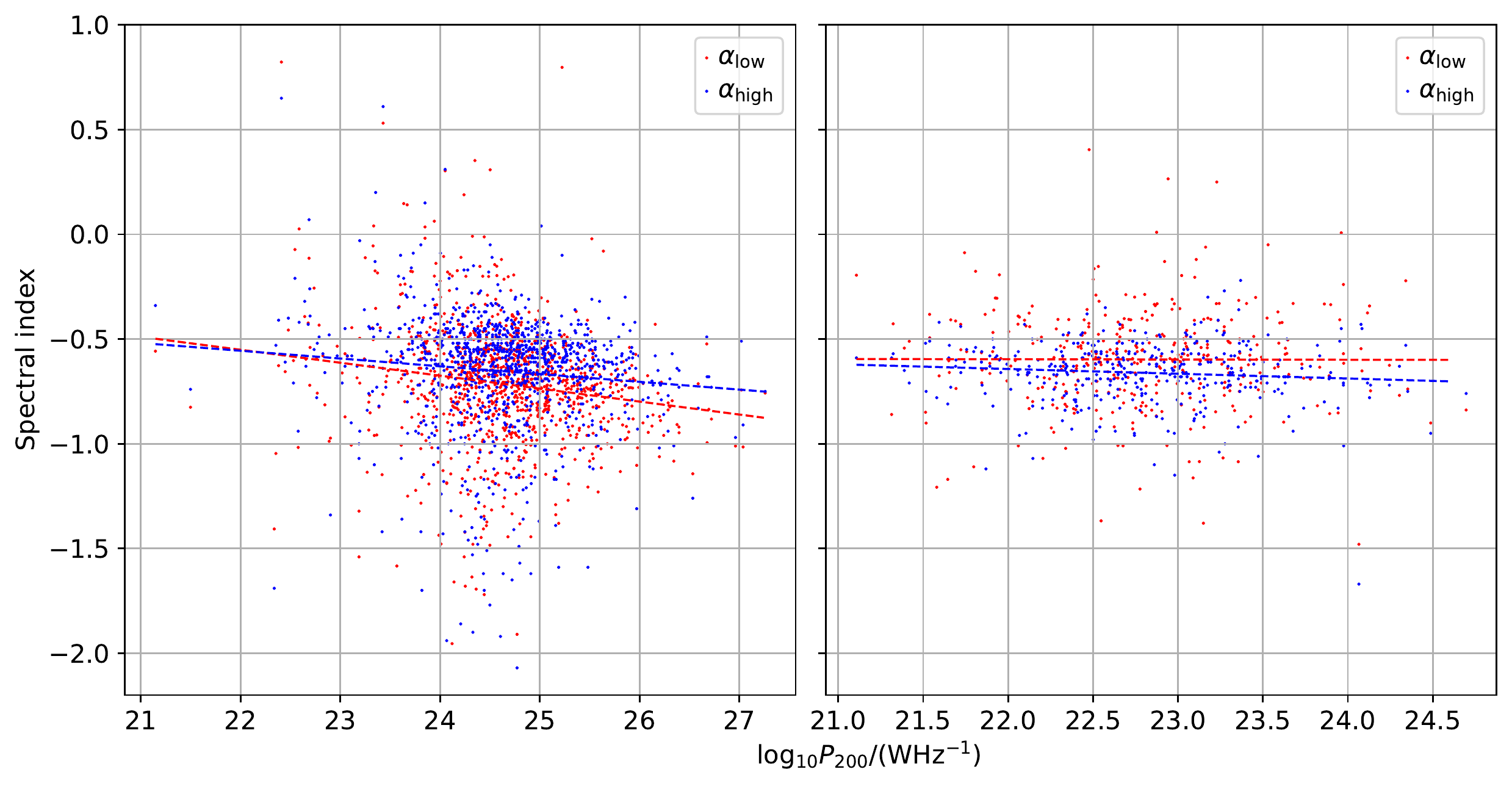}
\caption{$\alpha_{\mathrm{low}}$ (red) and $\alpha_{\mathrm{high}}$ (blue) as a function of the 200-MHz radio luminosity for the AGN (left) and SF galaxies (right) in the GLEAM-6dFGS sample with measured $\alpha_{\mathrm{low}}$ and $\alpha_{\mathrm{high}}$. The dashed lines are the linear regression lines.}
\label{fig:alpha_vs_luminosity}
\end{center}
\end{figure*}

For the G4Jy sample, which extends to much higher redshifts and luminosities than the GLEAM-6dFGS sample, the spectral indices over the same frequency ranges are steeper by $\sim 0.15$: the median value of $\alpha$ is $-0.740 \pm 0.012$ between 151 and 843~MHz, $-0.786 \pm 0.006$ between 151 and 1400~MHz, and $-0.829 \pm 0.006$ within the GLEAM band \citep{white2020b}. In Fig.~\ref{fig:alpha_vs_luminosity}, we investigate the dependence of $\alpha_{\mathrm{low}}$ and $\alpha_{\mathrm{high}}$ on $P_{200\,\mathrm{MHz}}$ within the GLEAM-6dFGS sample, separating between the AGN and SF galaxies. The results of a linear regression analysis to quantify any correlation between the spectral index and radio luminosity for both populations are given in Table~\ref{tab:lum_alpha_correlation}. For the AGN, there is a statistically significant anti-correlation between $\alpha_{\mathrm{low}}$ and $P_{200\,\mathrm{MHz}}$, and $\alpha_{\mathrm{high}}$ and $P_{200\,\mathrm{MHz}}$; $\alpha_{\mathrm{low}}$ and $\alpha_{\mathrm{high}}$ steepen by $0.062 \pm 0.011$ and $0.037 \pm 0.011$ per logarithmic luminosity decade respectively. For the SF galaxies, there is no statistically significant correlation between $\alpha_{\mathrm{low}}$ and $P_{200\,\mathrm{MHz}}$, and $\alpha_{\mathrm{high}}$ and $P_{200\,\mathrm{MHz}}$.

\begin{table*}
\small
\centering
\caption{Results of a linear regression analysis to investigate the dependence of $\alpha_{\mathrm{low}}$ and $\alpha_{\mathrm{high}}$ on $P_{200\,\mathrm{MHz}}$ for the AGN and SF galaxies in the GLEAM-6dFGS sample. $\rho$ is the correlation coefficient. The null hypothesis that the slope is zero is evaluated using a Wald test.}
\label{tab:lum_alpha_correlation}
\begin{tabular}{@{} c | c c c | c c c}
\hline
& \multicolumn{3}{c|}{$\alpha_{\mathrm{low}}$}  & \multicolumn{3}{c}{$\alpha_{\mathrm{high}}$} \\
Class & $\rho$ & Probability of & $\Delta \alpha$ per log. & $\rho$ & Probability of & $\Delta \alpha$ per log. \\
& & null hypothesis & luminosity decade & & null hypothesis & luminosity decade \\
\hline 
AGN & $-0.16$ & $\num{8.68E-08}$ & $-0.062\pm0.011$ & $-0.10$ & $\num{1.10E-03}$ & $-0.037\pm0.011$ \\ 
 SF & $-0.00$ & $\num{9.51E-01}$ & $-0.001\pm0.020$ & $-0.10$ & $\num{6.71E-02}$ & $-0.023\pm0.012$ \\
\hline
\end{tabular}
\end{table*}

\cite{laing1980} investigated the relation between spectral index and radio luminosity for a complete sample of 165 sources with $S_{178\,\mathrm{MHz}} > 10$~Jy, consisting primarily of FRII sources. For the extended structure in the FRII sources, $\alpha_{0.75\,\mathrm{GHz}}^{5\,\mathrm{GHz}}$ was found to be anti-correlated with the 1.4~GHz radio luminosity. The slope of the regression line (i.e. the change in the spectral index per logarithmic luminosity decade) was measured as --0.088. A similar anti-correlation was found between $\alpha_{0.75\,\mathrm{GHz}}^{5\,\mathrm{GHz}}$ and redshift, which is expected given the strong correlation between luminosity and redshift in bright, flux-limited samples. The $P-\alpha$ anti-correlation was confirmed by \cite{blundell1999} for a number of complete samples of FRII sources selected at frequencies close to 151~MHz. They suggested that the anti-correlation is caused by radiative losses in the enhanced magnetic fields of the hotspots of sources with more powerful jets.

We find that the $P-\alpha$ anti-correlation continues to hold for less powerful sources in the local universe. \cite{degasperin2018} investigated the spectral index properties between 147 and 1400~MHz of radio sources from TGSS and NVSS, and reported a flattening of the median spectral index with decreasing flux density. A similar spectral flattening towards lower flux densities has been observed in deeper radio samples covering smaller sky areas \citep[e.g.][]{prandoni2006,intema2011,whittam2013}. Since compact radio cores display much flatter spectra and are typically much fainter than the lobes of well-evolved radio galaxies, \citeauthor{degasperin2018} argued that this trend is driven by the presence of core-dominated and young AGN at low flux densities. Our results for the GLEAM-6dFGS AGN are consistent with this picture. Flat-spectrum sources are more common at lower radio luminosity, suggesting the existence of a significant population of low-luminosity AGN that remain core-dominated even at low frequencies.

\begin{figure*}
\begin{center}
\includegraphics[scale=0.65, trim=0cm 0cm 0cm 0cm]{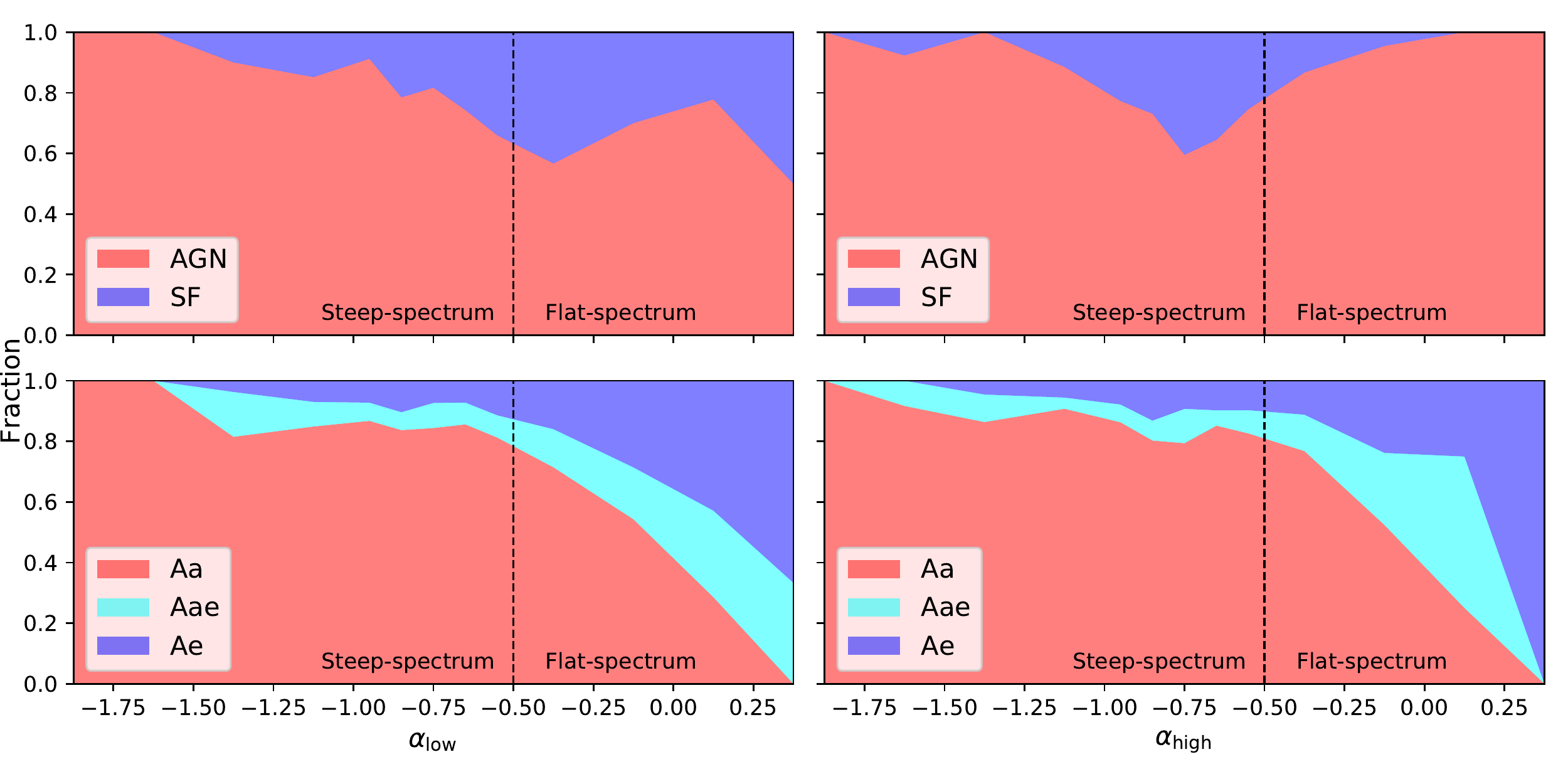}
\caption{Top: fractions of AGN and SF galaxies as a function of $\alpha_{\mathrm{low}}$ (left) and $\alpha_{\mathrm{high}}$ (right). Bottom: among the AGN, fractions of Aa, Aae and Ae sources as a function of $\alpha_{\mathrm{low}}$ (left) and $\alpha_{\mathrm{high}}$ (right).}
\label{fig:fractionAGN_vs_alpha}
\end{center}
\end{figure*}

Fig.~\ref{fig:fractionAGN_vs_alpha} shows that there is a strong increase in the fraction of sources with optical emission lines with $\alpha_{\mathrm{low}}$ and $\alpha_{\mathrm{high}}$. In other words, optical emission lines are much more common in galaxies which host flat-spectrum radio sources. In their study of the local radio source population at 20~GHz, \cite{sadler2013} observed the same trend for radio galaxies with weak optical emission lines (class `Aae') when measuring spectral indices between $\sim 1$ and 20~GHz. The fraction of radio galaxies with strong emission lines (class `Ae') was not found to change significantly with spectral index.

\cite{zajacek2019} studied radio spectral index trends of SDSS-FIRST sources across optical emission line diagnostic diagrams to search for potential correlations. They found a trend of decreasing 4.85--10.45~GHz spectral index with increasing ionisation ratio [OIII]/H$\beta$, which corresponds to the LINER-Seyfert transition in the optical diagnostic diagrams. They interpreted this trend as a result of re-triggered nuclear and jet activity. 

In low-frequency selected samples, there is preferential detection of sources with ultra-steep spectra down to low frequencies. A significant fraction of AGN in the GLEAM-6dFGS sample have ultra-steep spectra: 4.2 per cent have $\alpha_{\mathrm{low}} < -1.2$ and 4.6 per cent have $\alpha_{\mathrm{high}} < -1.2$.\cite{mahatma2018} reported a similar fraction (4.1 per cent) of sources with $\alpha_{150}^{1400} < -1.2$ in their LOFAR study of remnant radio-loud AGN in the Herschel-ATLAS field. The ultra-steep-spectrum AGN in our sample have a wide range of radio luminosities; the median value of $P_{200\,\mathrm{MHz}}$ for the AGN with $\alpha_{\mathrm{low}} < -1.2$ ($10^{24.41}~\mathrm{W}~\mathrm{Hz}^{-1}$) is slightly lower than the median value of $P_{200\,\mathrm{MHz}}$ for all AGN ($10^{24.61}~\mathrm{W}~\mathrm{Hz}^{-1}$). \cite{quici2021} show that remnant radio galaxies can be reliably identified by their ultra-steep spectra at low frequency ($\lesssim 400$~MHz), although the technique fails to capture younger remnants with steeper spectra at high frequencies. The absence of a radio core at high frequency is needed to establish whether these are genuine remnant radio galaxies.

\subsubsection{SF galaxies}\label{SF radio spectra}

For the SF galaxies, there is a slight flattening in the typical spectral index towards lower frequency: the median value of $\alpha_{\mathrm{low}} = -0.596 \pm 0.015$ and the median value of $\alpha_{\mathrm{high}} = -0.650 \pm 0.010$. A larger fraction of SF galaxies have flat spectra at low frequency: 10 per cent have $\alpha_{\mathrm{high}} > -0.5$ while 30 per cent have $\alpha_{\mathrm{low}} > -0.5$. We also find that the distribution of $\alpha_{\mathrm{low}}$ is significantly broader than that of $\alpha_{\mathrm{high}}$. We are able to show that this result is robust and not just a consequence of the larger measurement errors on $\alpha_{\mathrm{low}}$. Using Monte Carlo simulations which account for the measurement errors, the intrinsic dispersions in $\alpha_{\mathrm{low}}$ and $\alpha_{\mathrm{high}}$ are estimated to be $0.186\pm0.012$ and $0.127\pm0.008$ respectively. These simulations are described in detail in Appendix~\ref{Simulations to estimate the intrinsic spectral index dispersion for the SF galaxies}.

\cite{chyzy2018} also found a flattening in the typical spectral index towards lower frequency for a sample of 106 nearby SF galaxies using measurements over a larger frequency range. Using 150-MHz flux density measurements from the LOFAR Multifrequency Snapshot Sky Survey \citep[MSSS;][]{heald2015} and literature flux densities at various frequencies, they obtained a median value of $\alpha_{50\,\mathrm{MHz}}^{1.5\,\mathrm{GHz}} = -0.57$ and a median value of $\alpha_{1.3\,\mathrm{GHz}}^{5\,\mathrm{GHz}} = -0.77$. Since there was no tendency for the highly-inclined galaxies to have flatter low-frequency spectra, they argued that the observed flattening was not due to thermal absorption. From numerical modeling of the radio emission, they inferred that the flattening resulted principally from synchrotron spectral curvature due to cosmic ray energy losses and propagation effects.

\section{The local radio luminosity function at 200~MH\lowercase{z}}\label{The local radio luminosity function at 200 MHz}

\subsection{Calculating the local RLF}\label{Calculating the local RLF}

We calculate the local RLF using the $1/V_{\mathrm{max}}$ method of \cite{schmidt1968}, where $V_{\mathrm{max}}$ is the maximum volume within which a galaxy can satisfy all the sample selection criteria ($S_{200\,\mathrm{MHz}} > 100$~mJy in the shallow region, $S_{200\,\mathrm{MHz}} > 55$~mJy in the deep region, and $K \leq 12.65$~mag in both the shallow and deep regions). The luminosity function in a given luminosity bin centred at $L$ is given by 
\begin{equation}
\Phi(L) = \sum\limits_{i=1}^N \left( \frac{1}{V_{\mathrm{max,\,deep},\,i} + V_{\mathrm{max,\,shallow},\,i} } \right) \,,
\end{equation}
where $N$ is the number of galaxies in the luminosity bin, and $V_{\mathrm{max,\,deep},\,i}$ and $V_{\mathrm{max,\,shallow},\,i}$ are the maximum volumes within which the $i^{\mathrm{th}}$ galaxy can satisfy the sample selection criteria in the deep and shallow regions respectively. The rms Poisson counting error in $\Phi(L)$ is given by
\begin{equation}
d\Phi(L) = \left[ \sum\limits_{i=1}^N \left( \frac{1}{V_{\mathrm{max,\,deep},\,i} + V_{\mathrm{max,\,shallow},\,i} } \right)^2 \right]^{1/2} \,.
\end{equation}

We calculate the luminosity function in 13 luminosity bins of width 1~mag ($=0.4~\mathrm{dex}$) between $10^{21.6}$ and $10^{26.8}~\mathrm{W}~\mathrm{Hz}^{-1}$, and one luminosity bin of width 2~mag between $10^{26.8}$ and $10^{27.6}~\mathrm{W}~\mathrm{Hz}^{-1}$. In luminosity bins where the number, $N$, of galaxies is small ($N < 5$), we use 84 per cent confidence upper and lower limits based on Poisson statistics, tabulated in \cite{gehrels1986}.

Following \cite{condon2019}, we correct the measured luminosity function for galaxy clustering at a distance $s$ centred on our own Galaxy using
\begin{equation}
\frac{\Phi_{\mathrm{P}}}{\Phi} = 1 + \frac{3}{3 - \gamma_{\mathrm{s}}} \left( \frac{s_{0}}{s} \right)^{\gamma_{\mathrm{s}}}  \,,
\end{equation}
where $\frac{\Phi_{\mathrm{P}}}{\Phi}$ is the expected overdensity near our own Galaxy, or the space density, $\Phi_{\mathrm{P}}$, of local galaxies divided by the average space density, $\Phi$, of all galaxies, $\gamma_{\mathrm{s}} = 1.66$ and  $s_{0} = 3.76 h^{-1}$~Mpc \citep{fisher1994}. The volume within $s$ is multiplied by $\frac{\Phi_{\mathrm{P}}}{\Phi}$ when calculating $V_{\mathrm{max}}$; the correction is only applied for $s < 20 h^{-1}$~Mpc.

The completeness of the GLEAM-6dFGS cross-matched sample is estimated to be 95 per cent (see Section~\ref{Completeness and reliability of cross-matched sample}). To correct the luminosity function for incompleteness, we therefore multiply $\Phi(L)$ in all bins by $\frac{1}{0.95}$. Additionally, we need to account for spectroscopic incompleteness in the 6dFGS catalogue, i.e. the fact that some galaxies in the 6dFGS target list were either not able to be observed, or had only a poor-quality spectrum so that a reliable redshift could not be measured. The 6dFGS spectroscopic completeness varies across the sky, but is estimated to be 85 per cent for the NVSS/SUMSS-6dFGS sample as a whole. To correct the luminosity function for spectroscopic incompleteness, we therefore multiply $\Phi(L)$ in all bins by a further factor of $\frac{1}{0.85}$.

\subsection{Results}\label{Results}

\begin{figure}
\begin{center}
\includegraphics[scale=0.9, trim=0cm 0cm 0cm 0cm]{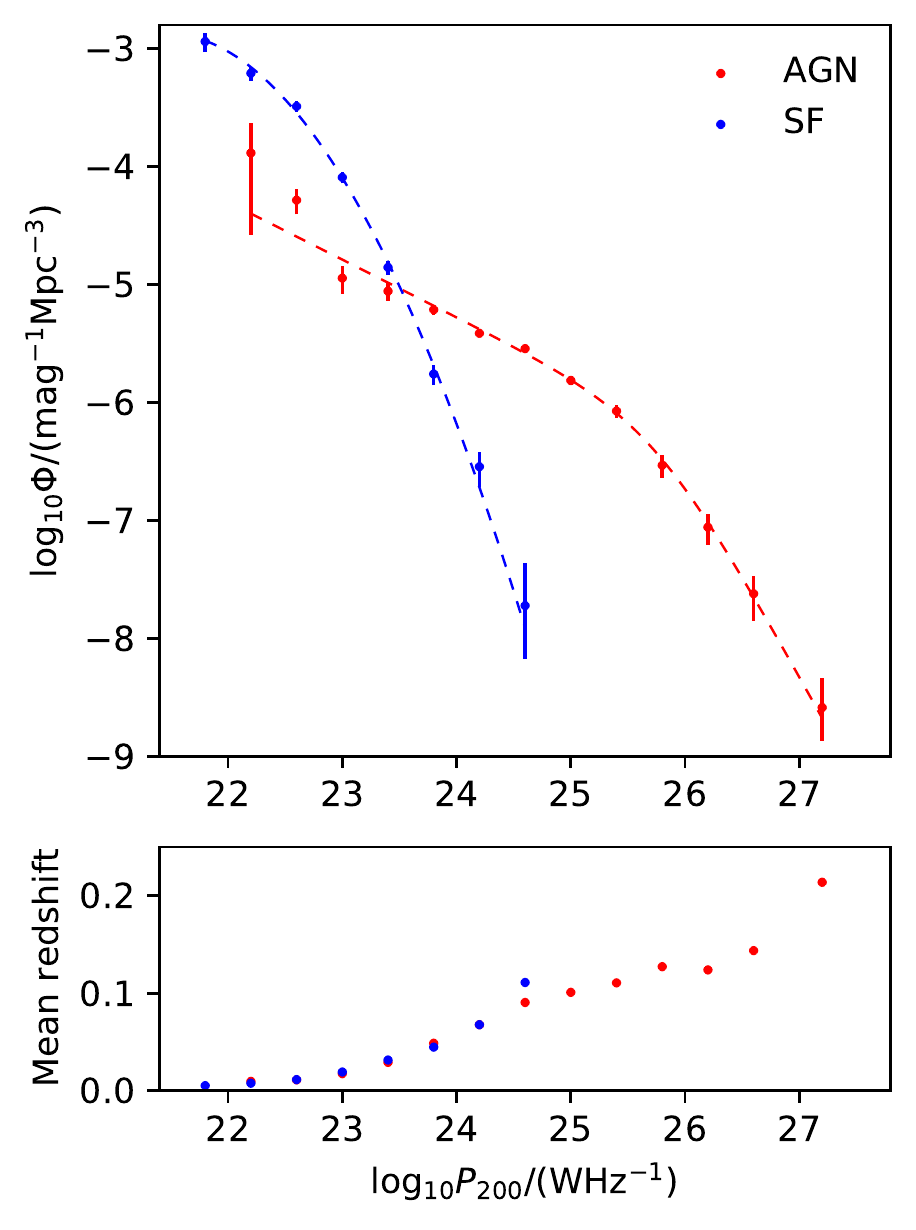}
\caption{Top: the local RLFs at 200~MHz for the AGN (red) and SF galaxies (blue) in the GLEAM-6dFGS sample. The two dashed curves correspond to the double power-law and \cite{saunders1990} fits to the AGN and SF data, respectively. Bottom: the mean redshifts of the AGN and SF galaxies in each bin.}
\label{fig:RLF_class}
\end{center}
\end{figure}

Our measured local RLF at 200~MHz for AGN and SF galaxies is tabulated in Table~\ref{tab:RLF} and plotted in the top panel of Fig.~\ref{fig:RLF_class}. The six sources with unknown optical spectra are classified according to their 200~MHz luminosity: those with $P_{200\,\mathrm{MHz}} > 10^{23.5}~\mathrm{W}~\mathrm{Hz}^{-1}$ are classed as AGN and those with $P_{200\,\mathrm{MHz}} \leq 10^{23.5}~\mathrm{W}~\mathrm{Hz}^{-1}$ are classed as SF galaxies. In total, 3 AGN and 44 SF galaxies are affected by the local volume correction described above. This correction only significantly affects the measured luminosity function of the SF galaxies in the lowest luminosity bin.

\begin{table*}
\small
\centering
\caption{Local RLFs at 200~MHz for all radio sources, the AGN and SF galaxies in the GLEAM-6dFGS sample.}
\label{tab:RLF}
\begin{tabular}{@{} c c c c c c c} 
\hline
& \multicolumn{2}{c}{All galaxies}
& \multicolumn{2}{c}{AGN}
& \multicolumn{2}{c}{SF galaxies}\\
$\log_{10} P_{200}$
& $N$
& $\log_{10} \Phi$
& $N$
& $\log_{10} \Phi$
& $N$
& $\log_{10} \Phi$ \\
(W~$\mathrm{Hz}^{-1}$)
& 
& ($\mathrm{mag}^{-1}~\mathrm{Mpc}^{-3}$)
& 
& ($\mathrm{mag}^{-1}~\mathrm{Mpc}^{-3}$)
& 
& ($\mathrm{mag}^{-1}~\mathrm{Mpc}^{-3}$) \\
\hline
21.8 & 32 & $-2.94^{+0.07}_{-0.09}$ & 0 &  & 32 & $-2.94^{+0.07}_{-0.09}$ \\
22.2 & 65 & $-3.13^{+0.07}_{-0.09}$ & 5 & $-3.88^{+0.25}_{-0.70}$ & 60 & $-3.21^{+0.06}_{-0.06}$ \\
22.6 & 131 & $-3.42^{+0.04}_{-0.04}$ & 20 & $-4.29^{+0.09}_{-0.12}$ & 111 & $-3.49^{+0.04}_{-0.05}$ \\
23.0 & 121 & $-4.04^{+0.04}_{-0.05}$ & 17 & $-4.95^{+0.10}_{-0.13}$ & 104 & $-4.09^{+0.04}_{-0.05}$ \\
23.4 & 106 & $-4.64^{+0.04}_{-0.05}$ & 41 & $-5.06^{+0.07}_{-0.08}$ & 65 & $-4.86^{+0.05}_{-0.06}$ \\
23.8 & 147 & $-5.11^{+0.04}_{-0.04}$ & 118 & $-5.21^{+0.04}_{-0.05}$ & 29 & $-5.76^{+0.08}_{-0.09}$ \\
24.2 & 233 & $-5.38^{+0.03}_{-0.03}$ & 221 & $-5.41^{+0.03}_{-0.03}$ & 12 & $-6.54^{+0.12}_{-0.17}$ \\
24.6 & 306 & $-5.54^{+0.03}_{-0.03}$ & 304 & $-5.54^{+0.03}_{-0.03}$ & 2 & $-7.72^{+0.36}_{-0.45}$ \\
25.0 & 211 & $-5.81^{+0.04}_{-0.04}$ & 211 & $-5.81^{+0.04}_{-0.04}$ & 0 &  \\
25.4 & 132 & $-6.07^{+0.05}_{-0.06}$ & 132 & $-6.07^{+0.05}_{-0.06}$ & 0 &  \\
25.8 & 60 & $-6.53^{+0.09}_{-0.11}$ & 60 & $-6.53^{+0.09}_{-0.11}$ & 0 &  \\
26.2 & 19 & $-7.06^{+0.11}_{-0.15}$ & 19 & $-7.06^{+0.11}_{-0.15}$ & 0 &  \\
26.6 & 8 & $-7.62^{+0.15}_{-0.23}$ & 8 & $-7.62^{+0.15}_{-0.23}$ & 0 &  \\
27.2 & 4 & $-8.59^{+0.25}_{-0.28}$ & 4 & $-8.59^{+0.25}_{-0.28}$ & 0 &  \\
Total & 1575 & & 1160 & & 415 & \\
\hline
\end{tabular}
\end{table*}

The RLF for the GLEAM-6dFGS AGN is well-fitted using a double power-law function of the form
\begin{equation}
\Phi(L) = \frac{C}{ (L/L_{\star})^\alpha + (L/L_{\star})^\beta } \,.
\label{eqn:double_power_law_function}
\end{equation}
The best-fitting parameters of equation~\ref{eqn:double_power_law_function} are $C = 10^{-6.13} \, \mathrm{mag}^{-1} \, \mathrm{Mpc}^{-3}$, $L_{\star} = 10^{25.76} \, \mathrm{W} \, \mathrm{Hz}^{-1}$, $\alpha = 1.76$ and  $\beta = 0.49$, where $\alpha$ is the power-law slope for $L \gg L_{\star}$ and $\beta$ is the power-law slope for $L \ll L_{\star}$. 

The RLF for SF galaxies is commonly fitted by the parametric form given by
\begin{equation}
\Phi(L) = C \left( \frac{L}{L_{\star}} \right)^{1-\alpha} \mathrm{exp} \left\{ -\frac{1}{2} \left[ \frac{\log_{10} (1+L/L_{\star})}{\sigma} \right]^{2} \right\} \,,
\label{eqn:saunders_function}
\end{equation}
analogous to the far-infrared luminosity function of \textit{Infrared Astronomical Satellite (IRAS)} galaxies \citep{saunders1990}. This function approaches a power-law at $L \ll L_{\star}$ and falls like a Gaussian at $L \gg L_{\star}$. The high luminosity end of the function is well sampled by the GLEAM-6dFGS SF galaxies. The best-fitting parameters of equation~\ref{eqn:saunders_function} are $C = 10^{-2.84} \, \mathrm{mag}^{-1} \, \mathrm{Mpc}^{-3}$, $L_{\star} = 10^{21.06} \, \mathrm{W} \, \mathrm{Hz}^{-1}$, $\alpha = 0.68$ and $\sigma = 0.66$.

While the AGN/SF galaxy classification from the optical spectra is generally reliable, there are a couple of cases in which an AGN can be mis-classified as a SF galaxy or vice-versa: firstly, the galaxy has a radio-quiet optical AGN but the radio emission comes mainly from star formation \citep[e.g.][]{kauffmann2003}. In nearby galaxies, the main SF regions may also fall outside the region covered by the optical fibre used for spectroscopy, aggravating this problem. Secondly, the optical spectrum is SF-like, but the radio emission comes mainly from a central AGN. Here, the AGN may be dust-obscured in the optical, as may be the case in ULIRGs \citep[e.g.][]{norris2013}, or else the radio emission could be a genuine composite of AGN and SF emission. Since SF galaxies outnumber AGN at $P_{200\,\mathrm{MHz}} \lesssim 10^{23.5}~\mathrm{W}~\mathrm{Hz}^{-1}$, it is more likely that a genuine SF object is mis-classified as an AGN rather than the other way around. This could potentially account for part of the AGN upturn seen in the two lowest-luminosity bins in Fig.~\ref{fig:RLF_class}.

The bottom panel of Fig.~\ref{fig:RLF_class} shows the mean redshifts of the AGN and SF galaxies in each luminosity bin. As expected, the mean redshift increases with luminosity due to the Malmquist bias. We note that the small change in the redshift distribution with luminosity will affect the shape of the luminosity function due to redshift evolution of the AGN and star-forming populations.

\subsection{Comparison of recent RLF measurements}\label{Comparison of recent RLF measurements}

\begin{figure*}
\begin{center}
\includegraphics[scale=0.65, trim=0cm 0cm 0cm 0cm]{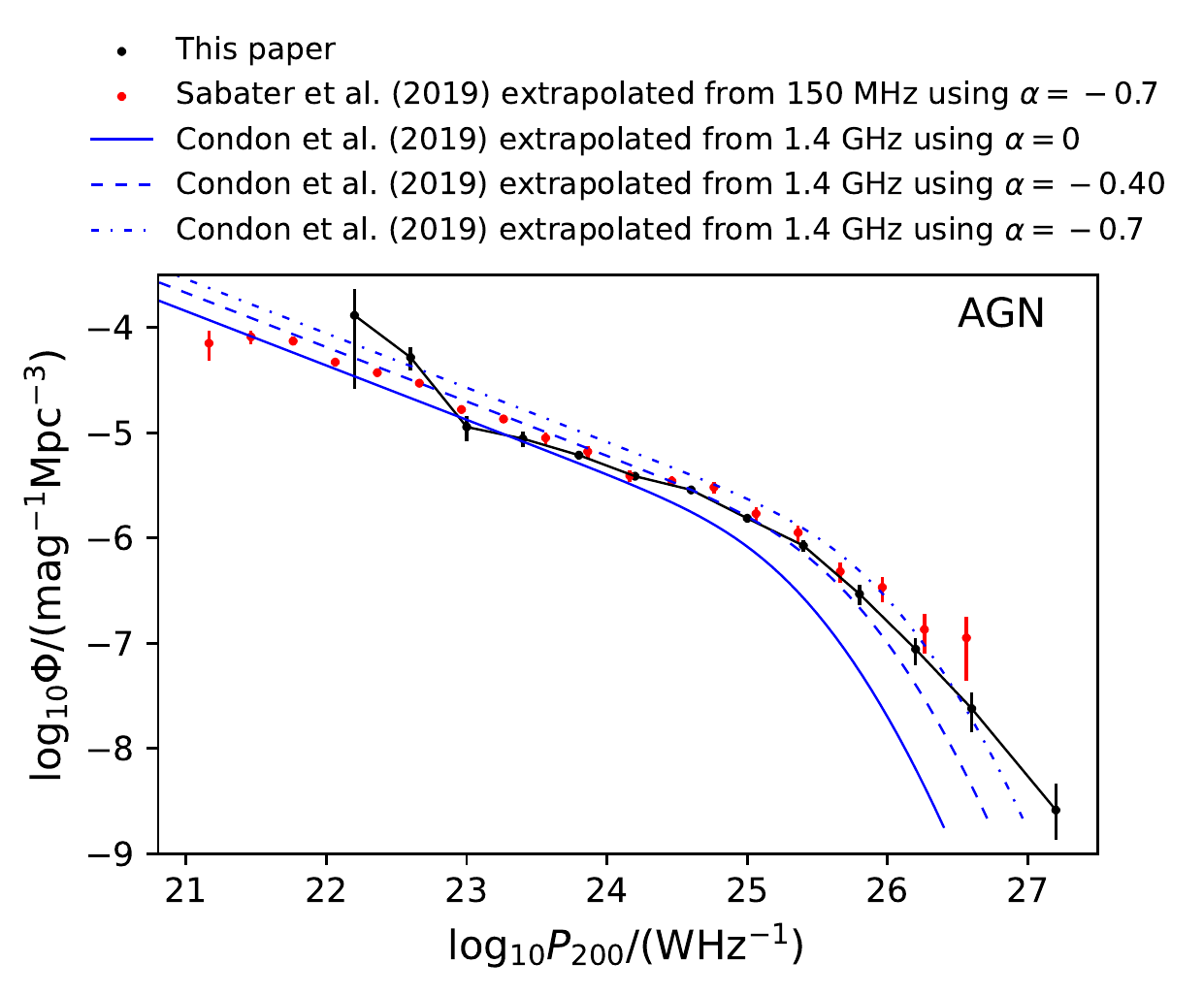}
\includegraphics[scale=0.65, trim=0cm 0cm 0cm 0cm]{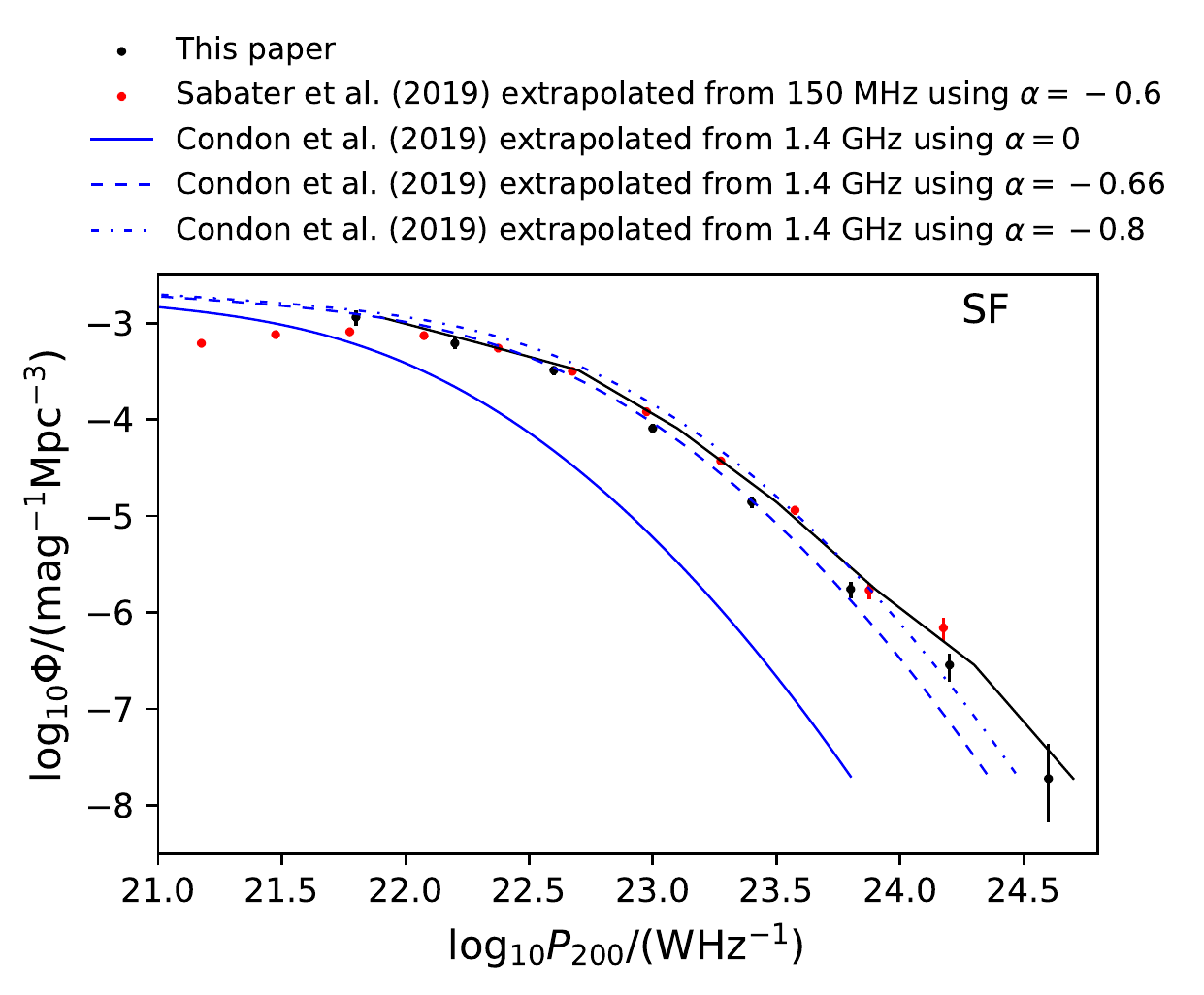}
\caption{Left: the local RLF at 200~MHz for AGN measured in this paper (black circles). The local RLF at 150~MHz for AGN from \citet{sabater2019} extrapolated to 200~MHz assuming a spectral index of --0.7 (red circles). The local RLF at 1400~MHz for AGN using the parameterisation of \citet{condon2019}, extrapolated to 200~MHz assuming spectral indices of 0 (solid blue line), --0.40 (dashed blue line) and --0.7 (dot-dashed blue line), as discussed in the text. Right: the local RLF at 200~MHz for SF galaxies measured in this paper (black circles). The local RLF at 150~MHz for SF galaxies from \citeauthor{sabater2019} extrapolated to 200~MHz assuming a spectral index of --0.6 (red circles). The local RLF at 1400~MHz for SF galaxies using the parameterisation of \citet{condon2019}, extrapolated to 200~MHz assuming spectral indices of 0 (solid blue line), --0.66 (dashed blue line) and --0.8 (dot-dashed blue line), as discussed in the text.}
\label{fig:RLF_comparison}
\end{center}
\end{figure*}

In the left panel of Fig.~\ref{fig:RLF_comparison}, we compare the local RLF for AGN measured in this paper with the local RLF at 150~MHz for AGN from the LoTSS-SDSS sample \citep{sabater2019}, extrapolated to 200~MHz assuming $\alpha = -0.7$, the typical spectral index at low frequency of AGN measured in this paper. Overall, the luminosity functions are in good agreement. The median redshift (0.143) of the AGN in the LoTSS-SDSS sample is higher than that (0.081) in the GLEAM-6dFGS sample. Cosmological evolution of HERGs, which dominate the local AGN population at $P_{1.4\,\mathrm{GHz}} \gtrsim 10^{26}~\mathrm{W}~\mathrm{Hz}^{-1}$ \citep[see e.g.][]{heckman2014,pracy2016}, may therefore partly explain the slight offset between the luminosity functions at the high-luminosity end.

We use the parameterised form of the 1400~MHz local RLF for AGN from equation~28 of \cite{condon2019} and fit this to the 200~MHz data by making a shift in radio power set by a single characteristic 200--1400~MHz spectral index. The best-fitting value of $\alpha$ is --0.40, which is flatter than the median value of $\alpha_{\mathrm{high}}$ (--0.60) for the AGN in the GLEAM-6dFGS sample. The AGN sample by \cite{condon2019} has a slightly higher median redshift (0.12). Any luminosity evolution of the AGN population would therefore cause the best-fitting value of $\alpha$ to be flatter than expected. We note, however, that the local AGN population is dominated by LERGs at $P_{1.4\,\mathrm{GHz}} \lesssim 10^{26}~\mathrm{W}~\mathrm{Hz}^{-1}$, which display little or no evolution \citep[see e.g.][]{heckman2014,pracy2016}. There is a significant change in the shape of the luminosity function. At lower luminosity, a flatter spectral index is needed to obtain a good fit. This is consistent with our earlier investigation of spectral index versus luminosity in Section~\ref{Radio spectra}, where we found that the typical value of $\alpha_{\mathrm{high}}$ flattens with decreasing luminosity.

\begin{figure}
\begin{center}
\includegraphics[scale=0.55, trim=0cm 0cm 0cm 0cm]{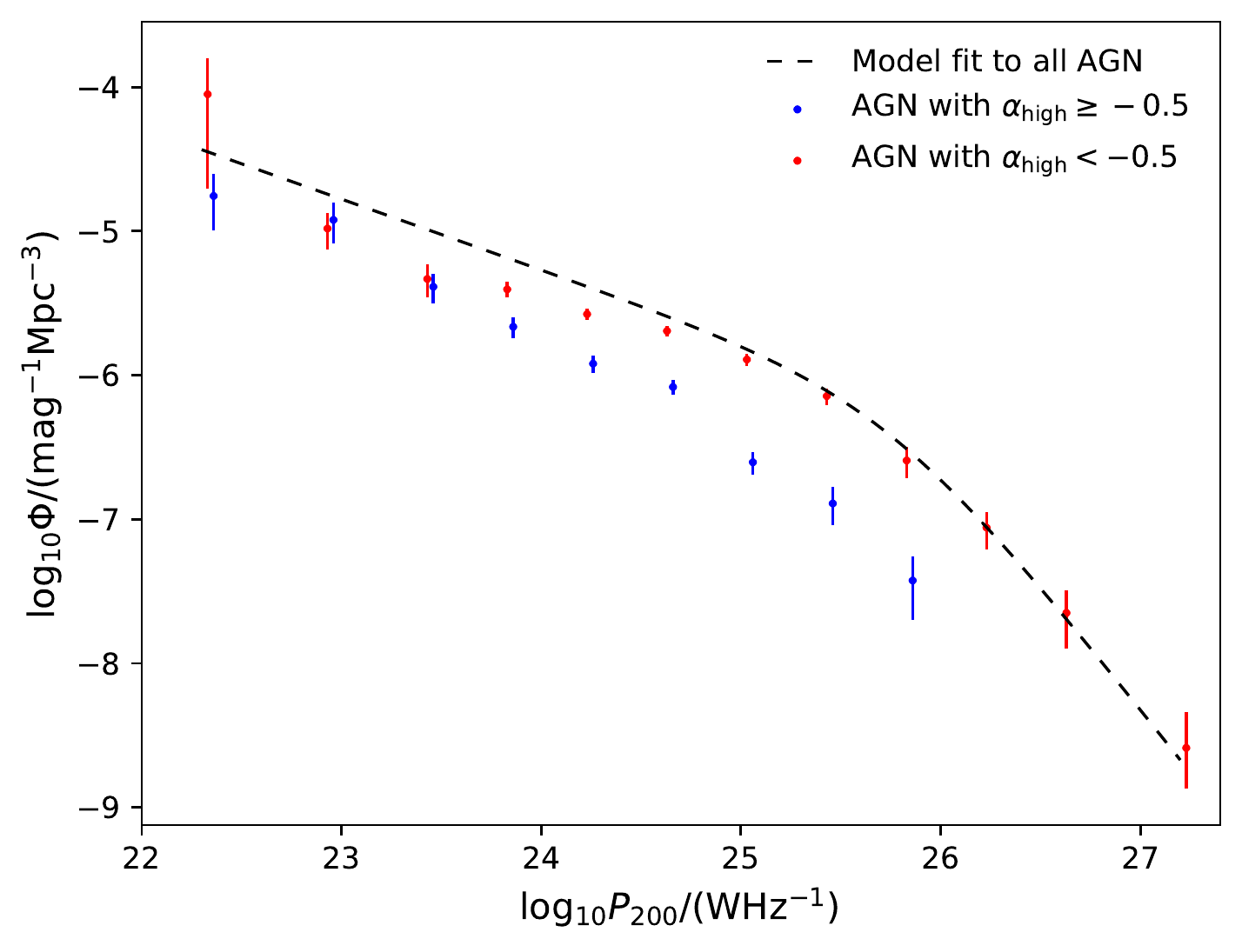}
\caption{The local RLF for the GLEAM-6dFGS AGN, separating between flat-spectrum (blue points) and steep-spectrum (red points) sources. The dashed curve shows the double power-law fit to the local RLF for all AGN. There is an increase in the relative contribution of flat-spectrum sources with decreasing luminosity.}
\label{fig:RLF_alpha}
\end{center}
\end{figure}

In Fig.~\ref{fig:RLF_alpha}, we measure the local RLF for the AGN in the GLEAM-6dFGS sample, separating between flat-spectrum ($\alpha_{\mathrm{high}} > -0.5$) and steep-spectrum ($\alpha_{\mathrm{high}} < -0.5$) sources. While steep-spectrum sources overwhelmingly dominate the luminosity function at the highest luminosities, the relative contribution to the luminosity function from flat-spectrum sources gradually increases with decreasing luminosity.

In the right panel of Fig.~\ref{fig:RLF_comparison}, we compare the local RLF for SF galaxies measured in this paper with the local RLF at 150~MHz for SF galaxies by \cite{sabater2019}, extrapolated to 200~MHz assuming $\alpha = -0.6$, the typical spectral index at low frequency of SF galaxies measured in this paper. The luminosity function by \citeauthor{sabater2019} is shifted by $\log_{10} P_{200} \approx 0.1$ to the right relative to our measurements. As mentioned in Section~\ref{Introduction}, the LoTSS-SDSS sample covers a much smaller area of sky ($424~\mathrm{deg}^{2}$) and extends to much lower flux densities ($\approx 0.4$~mJy) than the GLEAM-6dFGS sample. Consequently, the median redshift (0.097) of the SF galaxies in the LoTSS-SDSS sample is significantly higher than that (0.015) in the GLEAM-6dFGS sample. \cite{ocran2020} studied the properties of SF galaxies selected at 610~MHz with the GMRT in a deep survey covering $\approx 1.86~\mathrm{deg}^{2}$. They explored the evolution of the RLF for SF galaxies out to $z \approx 1.5$. The SF galaxies were found to evolve as $f(z) = (1+z)^{(2.95 \pm 0.19)-(0.50 \pm 0.15)z}$ assuming pure luminosity evolution. The luminosity function by \citeauthor{sabater2019} is therefore expected to be shifted to the right by $\log_{10} P_{200} \approx \log_{10} [ f(0.097)/f(0.015) ] = 0.10$ relative to our measurements. We conclude that the luminosity functions are in good agreement after taking the rapid evolution of the SF galaxy population into account.

We use the parameterised form of the 1400~MHz local RLF for SF galaxies from equation~28 of \cite{condon2019} and fit this to the 200~MHz data using a single characteristic 200--1400~MHz spectral index. The median redshift of the SF galaxies in the NVSS-2MASX sample by \citeauthor{condon2019} is 0.06. The best-fitting value of $\alpha$ is --0.66, which is very close to the median value of $\alpha_{\mathrm{high}}$ (--0.65) for the SF galaxies in our sample.

\section{Summary and future work}\label{Summary and future work}

We have studied the local radio source population at 200~MHz over most of the southern sky ($16,679~\mathrm{deg}^{2}$) by combining the GLEAM Exgal and SGP data with optical spectroscopy from the 6dFGS DR3. The GLEAM-6dFGS sample contains 1,590 sources with a median redshift of 0.064 and 200~MHz radio luminosities in the range $\sim 10^{22}-10^{27}~\mathrm{W}~\mathrm{Hz}^{-1}$. The optical spectra indicate that 73 per cent of the sources are fuelled by AGN activity and 27 per cent by star formation. The vast majority (81 per cent) of the AGN have pure absorption-line spectra typical of giant elliptical galaxies. The AGN with emission-line spectra typically have lower luminosities and overlap in luminosity with the SF galaxies.

We characterise the typical spectra of the AGN and SF galaxies using the intra-band GLEAM spectral indices between 76 and 227~MHz, and spectral indices between 200~MHz and $\sim 1$~GHz measured by cross-matching the sample with NVSS and SUMSS. The relatively low spatial resolution of GLEAM, NVSS and SUMSS results in a high surface brightness sensitivity that is required to recover the extended emission in nearby galaxies and obtain accurate spectral indices.

For the AGN, the median value of $\alpha_{\mathrm{high}}$ ($-0.600 \pm 0.010$) is significantly flatter than the median value of $\alpha_{\mathrm{low}}$ ($-0.704 \pm 0.011$). This is likely the result of the cores of FRI sources becoming increasingly dominant at high frequencies. The 200~MHz luminosity is anti-correlated with $\alpha_{\mathrm{low}}$ and $\alpha_{\mathrm{high}}$; both $\alpha_{\mathrm{low}}$ and $\alpha_{\mathrm{high}}$ steepen by $\approx 0.05$ per logarithmic luminosity decade. A significant fraction ($\approx 4$ per cent) of the AGN in our low-frequency selected sample have ultra-steep spectra at low frequency with $\alpha_{\mathrm{low}} < -1.2$. These sources span a wide range of radio luminosities. Further work is needed to establish their nature. Optical emission lines are much more common in galaxies which host flat-spectrum radio sources, confirming the trend observed by \cite{sadler2013} at higher frequency for radio galaxies in the AT20G-6dFGS sample.

For the SF galaxies, the median value of $\alpha_{\mathrm{low}}$ is $\alpha = -0.596 \pm 0.015$ and the median value of $\alpha_{\mathrm{high}}$ is $\alpha = -0.650 \pm 0.010$. A larger fraction of SF galaxies have flat spectra at low frequency: 10 per cent have $\alpha_{\mathrm{high}} > -0.5$ while 30 per cent have $\alpha_{\mathrm{low}} > -0.5$. Unlike for the AGN population, there is no statistically significant correlation between the 200~MHz luminosity and $\alpha_{\mathrm{low}}$ and $\alpha_{\mathrm{high}}$. Further work is needed to identify the physical processes responsible for the spectral flattening at lower frequencies.

We derive the local RLF for AGN and SF galaxies at 200~MHz. A spectral index of --0.40 provides the best match between our local RLF for AGN and that measured by \cite{condon2019} at 1.4~GHz. This is significantly flatter than the median value of $\alpha_{\mathrm{high}}$ for the AGN in the GLEAM-6dFGS sample. The difference is likely due to the changing nature of the sources contributing to the local RLF at 200 and 1400~MHz, and the preferential selection of sources with flatter spectra at 1400~MHz. There appears to be a slight change in the shape of the local RLF for AGN with frequency due to an increase in the relative contribution of flat-spectrum sources ($\alpha_{\mathrm{high}} > -0.5$) at lower luminosity, although this effect is difficult to measure in the lowest one or two luminosity bins due to a potential misclassification of SF galaxies as AGN. A spectral index of --0.66 provides the best match between our local RLF for SF galaxies and that measured by \citeauthor{condon2019} at 1.4~GHz, which is very close to the median value of $\alpha_{\mathrm{high}}$ for the SF galaxies in the GLEAM-6dFGS sample.

Seymour et al., in preparation, are measuring the RLF at 200~MHz out to $z \approx 0.5$ by combining GLEAM SGP data with deep multi-wavelength surveys in CDFS and ELAIS-S1. The local RLF presented in this paper will provide an essential benchmark from which to analyse the cosmic evolution of radio galaxies to this redshift.

\begin{acknowledgements}
This scientific work makes use of the Murchison Radio-astronomy Observatory, operated by CSIRO. We acknowledge the Wajarri Yamatji people as the traditional owners of the Observatory site. Support for the operation of the MWA is provided by the Australian Government (NCRIS), under a contract to Curtin University administered by Astronomy Australia Limited. We thank the anonymous referee for helpful comments, which have substantially improved this paper. We acknowledge the Pawsey Supercomputing Centre which is supported by the Western Australian and Australian Governments. This research made use of the cross-match service provided by CDS, Strasbourg. CAJ thanks the Department of Science, Office of Premier \& Cabinet, WA for their support through the Western Australian Fellowship Program. 
Part of this research was supported by the Australian Research Council Centre of Excellence for All Sky Astrophysics (CAASTRO), through grant CE110001020.
\end{acknowledgements}

\section{Conflicts of Interest}
None.

\begin{appendix}

\section{Simulations to estimate the intrinsic spectral index dispersion for the SF galaxies}\label{Simulations to estimate the intrinsic spectral index dispersion for the SF galaxies}

As part of the spectral index analysis in Section~\ref{Radio spectra}, we perform Monte Carlo simulations to estimate the spectral index variability intrinsic to the SF galaxies taking into account the measurement errors. Given some intrinsic spectral index distribution, we predict the spectral index distribution that would be observed in the presence of measurement errors. We model the intrinsic spectral index distribution as a Gaussian\footnote{It appears reasonable to assume that the intrinsic spectral index distribution has a Gaussian form: the measured distributions of $\alpha_{\mathrm{low}}$ and $\alpha_{\mathrm{high}}$ for the SF galaxies can be well approximated by Gaussians, and the widths of these distributions are substantially broader than what can be explained by measurement errors alone.} centred on zero with standard deviation, $\sigma_{\mathrm{int}}$. We randomly draw the spectral indices of the 369 SF galaxies included in the spectral index analysis; the spectral index of the $i^{\mathrm{th}}$ source is given by
\begin{equation}
\alpha_{i} = \mathcal{N}(0,\, \sigma_{\mathrm{int}} ) + \mathcal{N}(0,\, \Delta \alpha_{i} ) \,,
\end{equation}
where $\mathcal{N}(0,\, \sigma_{\mathrm{int}} )$ denotes a random number drawn from a Gaussian distribution of mean 0 and standard deviation $\sigma_{\mathrm{int}}$, and $\Delta \alpha_{i}$ is the error on the measured spectral index ($\alpha_{\mathrm{low}}$ or $\alpha_{\mathrm{high}}$) of the $i^{\mathrm{th}}$ source taken from the GLEAM-6dFGS catalogue.

\begin{figure}
\includegraphics[scale=0.75, trim=0cm 0cm 0cm 0cm]{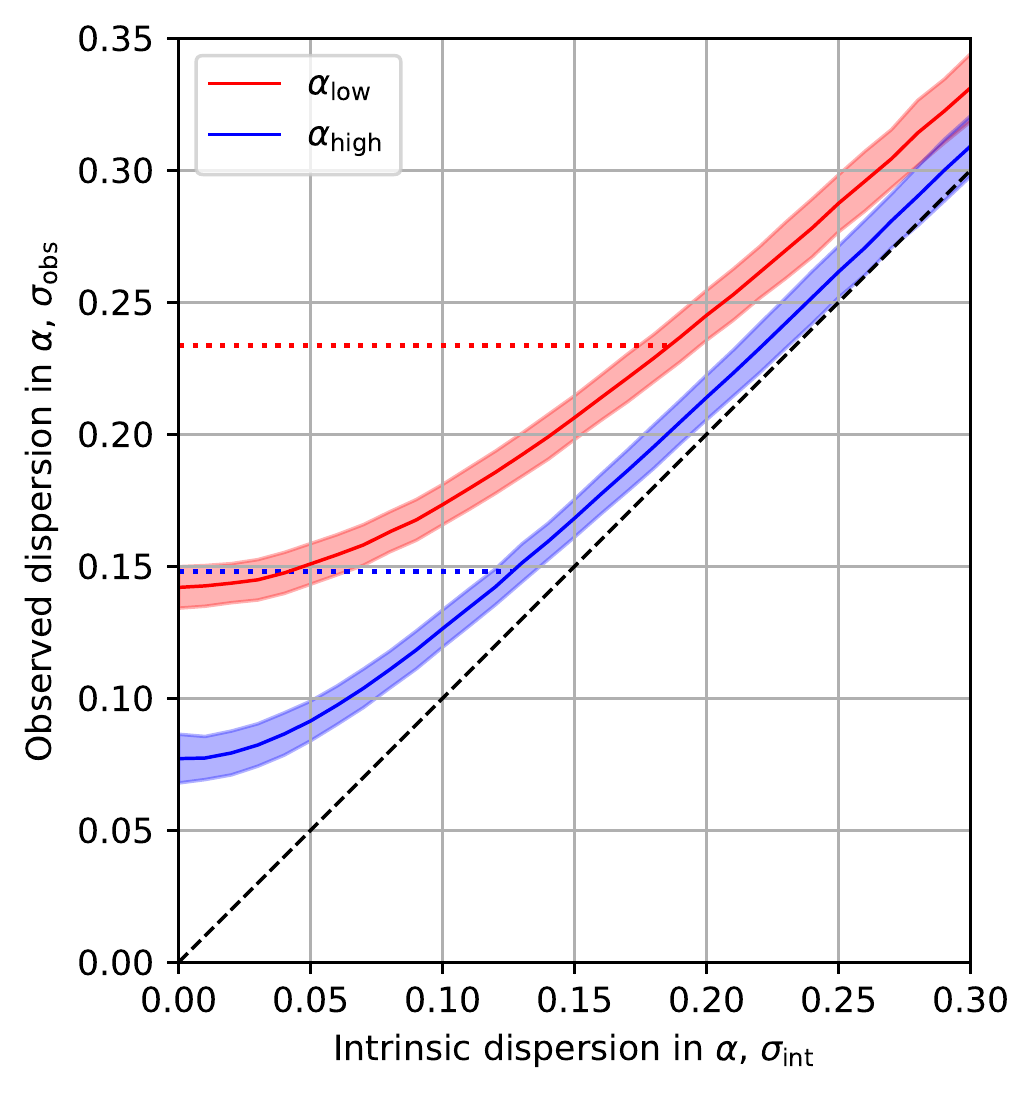}
\caption{Results of Monte Carlo simulations to estimate the variability in $\alpha_{\mathrm{low}}$ and $\alpha_{\mathrm{high}}$ intrinsic to the SF galaxies in the GLEAM-6dFGS sample. The dispersion in the spectral index that would be observed given the measurement errors is plotted as a function of the intrinsic spectral index dispersion; the shaded areas show the 1$\sigma$ errors. The dashed line indicates equal values of $\sigma_{\mathrm{obs}}$ and $\sigma_{\mathrm{int}}$. The dotted horizontal lines show the standard deviations of $\alpha_{\mathrm{low}}$ (0.233) and $\alpha_{\mathrm{high}}$ (0.148) for the SF galaxies in the real data.}
\label{fig:sim_alpha_errors}
\end{figure}

We then measure the standard deviation of the observed spectral index distribution, $\sigma_{\mathrm{obs}}$, as a function of $\sigma_{\mathrm{int}}$. We repeat the simulations 1,000 times to improve statistics. Fig.~\ref{fig:sim_alpha_errors} shows the mean value of $\sigma_{\mathrm{obs}}$ as a function of $\sigma_{\mathrm{int}}$ for $\alpha_{\mathrm{low}}$ (red curve) and $\alpha_{\mathrm{high}}$ (blue curve). The observed dispersion in $\alpha_{\mathrm{low}}$ is higher than that in $\alpha_{\mathrm{high}}$ due to the larger measurement errors on $\alpha_{\mathrm{low}}$. The two curves converge at high $\sigma_{\mathrm{int}}$ as the measurement errors on $\alpha_{\mathrm{low}}$ and $\alpha_{\mathrm{high}}$ become negligible with respect to $\sigma_{\mathrm{int}}$.

In the real data, the standard deviations of $\alpha_{\mathrm{low}}$ and $\alpha_{\mathrm{high}}$ are 0.233 and 0.148 respectively. From Fig.~\ref{fig:sim_alpha_errors}, we estimate the intrinsic dispersions in $\alpha_{\mathrm{low}}$ and $\alpha_{\mathrm{high}}$ to be  $0.186 \pm 0.012$ and $0.127 \pm 0.008$ respectively.

\end{appendix}

\bibliographystyle{pasa-mnras}
\bibliography{myreferences}
\setlength{\labelwidth}{0pt}

\label{lastpage}
\end{document}